\documentclass[reprint,
superscriptaddress,
 amsmath,amssymb,
 aps,
prl,
]{revtex4-2}

\usepackage{caption}
\usepackage{subcaption}
\usepackage{graphicx}% Include figure files
\usepackage{dcolumn}% Align table columns on decimal point
\usepackage[monochrome]{xcolor}
\usepackage{bm}% bold math
\usepackage{soul}
\usepackage{amsfonts,dsfont,xspace,relsize,mathtools,amsthm}% bm/graphicx/soul/amsmath/amssymb already loaded
\usepackage{bbm}% bold math
\usepackage{physics}
\usepackage{ulem}
\usepackage{hyperref}
\hypersetup{
	colorlinks = true,
	urlcolor   = blue,
	linkcolor  = blue,
	citecolor  = red
}
\usepackage{siunitx}
\usepackage[shortlabels]{enumitem}

\usepackage{tikz}

\newtheorem{theorem}{Theorem}
\newtheorem{corollary}{Corollary}
\newtheorem{lemma}{Lemma}
\newtheorem{definition}{Definition}
\newtheorem{proposition}{Proposition}
\newtheorem{remark}{Remark}
\newcommand{\eqdef}{:=}
\newcommand{\Ntri}{{\mathcal{N}}}
\newcommand{\Tz}{T_{0}}
\newcommand{\muinf}{\mu_\infty}
\newcommand{\muadv}{\mu_{\rm adv}}
\newcommand{\dstar}{d^{*}}

\DeclareMathOperator{\Var}{Var}

\newcommand{\1}{\mathbbm{1}}

\newcommand{\C}{\mathbb{C}}
\newcommand{\R}{\mathbb{R}}

\newcommand{\bxi}{{\boldsymbol \xi}}
\newcommand{\bx}{{\mathbf x}}
\newcommand{\bc}{{\mathbf c}}
\newcommand{\bv}{{\mathbf v}}
\newcommand{\bw}{{\mathbf w}}
\newcommand{\bk}{{\mathbf k}}
\newcommand{\bq}{{\mathbf q}}
\newcommand{\bz}{{\mathbf z}}
\newcommand{\cV}{{\mathcal V}}
\newcommand{\cA}{{\mathcal{A}}}

\newcommand{\cN}{{\mathcal{N}}}

\newcommand{\cP}{{\mathcal{P}}}

\newcommand{\cL}{{\mathcal{L}}}

\newcommand{\X}{{\mathsf{X}}}

\newcommand{\ii}{{\rm i}}
\makeatletter
\let\X\@undefined
\makeatother
\newcommand{\X}{{\mathcal{X}}}

\newcommand{\dketbra}[1]{|{#1}\rangle\langle{#1}|}

\newcommand{\Ord}[1]{\mathcal{O}\left( #1 \right)}

\newcommand{\ssection}[1]{\textit{#1} --}

\begin{document}

%\title{Quantum Self-Attention by Overlap Interference: \\ Predicting Classical and Many-Body Quantum Sequences}
%\title{Predicting Classical and Many-Body Quantum Sequences \\via Quantum Self-Attention}
%\title{Quantum Attention by Overlap Interference: \\Predicting Sequences from Classical and Many-Body Quantum Data}
\title{\color{red} Quantum Attention by Overlap Interference:\\ Predicting Classical and Many-Body Quantum Sequences}
\author{Alessio Pecilli}
\affiliation{Dipartimento di Ingegneria Civile, Informatica e delle Tecnologie Aeronautiche, Universit\`a degli Studi Roma Tre, Via della Vasca Navale 79, 00146 Rome, Italy}

\author{Matteo Rosati}
\affiliation{Dipartimento di Ingegneria Civile, Informatica e delle Tecnologie Aeronautiche, Universit\`a degli Studi Roma Tre, Via della Vasca Navale 79, 00146 Rome, Italy}

\begin{abstract}
We propose a variational quantum implementation of self-attention (QSA)—the core operation in transformers and large language models—which predicts future elements of a sequence by forming overlap-weighted combinations of past data. At variance with previous approaches, our QSA realizes the required nonlinearity through interference of state overlaps and a degree-$k$ polynomial kernel, and estimates a loss based on Rényi-$1/2$ entropic functionals via two observables' expectation values, avoiding the decoding of amplitude-encoded predictions into classical probabilities. QSA also accommodates a constrained, trainable data-embedding tying state overlaps to data-level similarities. Its dominant end-to-end training complexity scales as $O\left(\mu^{-1}k^2Td\right)$, versus $O\left(T d^{k+1}\right)$ of the fairest classical comparison, with $\mu$ a training signal; we show numerically that this allows a complexity advantage in the regime where sequence length $T$ dominates the embedding size $d$. In simulations, our QSA-based quantum transformer learns sequence prediction on classical data and on many-body transverse-field Ising trajectories—establishing trainable attention as a practical primitive for quantum dynamical modeling.
\end{abstract}

\maketitle

%\tableofcontents

\ssection{Introduction} 
Large language models, organized around transformer architectures, are rapidly revolutionizing human activities at large~\cite{Calvino2025}. A transformer takes sequences of data, embedded into lower-dimensional \textit{tokens}, and predicts the next datum in the sequence. At its heart lies the self-attention (SA) mechanism, which compares tokens through pairwise similarities, the {\it attention weights}, and uses these to build collective, context-dependent predictions; SA is also the dominant computational bottleneck, with a characteristic quadratic dependence on sequence length~\cite{Vaswani}.  

This has motivated a growing effort to “quantize” transformer components, aiming at asymptotic speedups or quantum-enhanced expressivity~\cite{Cherrat2022,Gao2023,Zhao2024,Zhao2024a,Liao2024,Guo2024,Khatri2024a}: quantization tailored to vision transformers~\cite{Cherrat2022}; Grover- or kernel-inspired constructions~\cite{Gao2023,Zhao2024,Zhao2024a}, some under structural assumptions on the attention matrix; faithful end-to-end mappings of transformer blocks onto circuits~\cite{Liao2024}; approximating or replacing attention with quantum singular value transform primitives~\cite{Guo2024,Khatri2024a}.

A persistent obstacle is that many such constructions store intermediate or final predictions in the amplitudes of a quantum state. While compact, this bottlenecks training: evaluating a classical loss can require additional assumptions, costly readout, or substantial classical post-processing, limiting end-to-end demonstrations. The data-to-token embedding is also often left implicit and fixed in advance; yet, since attention weights are built from token inner products, it determines which data-level relations are preserved and is thus central to any faithful quantization of SA.

Here we introduce a variational quantum self-attention (QSA) mechanism that harnesses the inner-product structure of classical self-attention (CSA) and directly outputs a loss value for each data sequence, enabling efficient training. Our QSA rests on three insights: (i) the SA output is an intrinsically non-linear function of token inner products, realizable by a suitable non-linear encoding of tokens into quantum states {\color{red} and a polynomial attention kernel, replacing softmax SA}; (ii) {\color{red} a loss based on Rényi-$1/2$ divergence can be calculated via the expectation of two observables on the circuit}, providing a hardware-friendly training signal; (iii) a constrained yet trainable embedding links token- to data-affinities. The resulting architecture inherits CSA's variational structure and {\color{red} approximates standard SA without long-term subroutines, being suitable for near/mid-term quantum hardware.

We prove that our QSA surpasses its CSA baseline in end-to-end training complexity, provided the training signal scales with a suitable negative power of the embedding size, and show by simulation that this advantage regime is attainable in the practically relevant limit of long training sequences. We then benchmark our simulated QSA against several CSA models, from baseline polynomial attention to standard softmax SA, finding that QSA reaches comparable loss values} in two generative modeling tasks: next-token prediction of classical sequences, and prediction of quantum-state sequences generated by many-body Hamiltonian evolution. The latter demonstrates, for the first time to our knowledge, that attention-like mechanisms can be trained as primitives for modeling quantum dynamical sequences, linking Hamiltonian simulation to quantum machine learning~~\cite{Puig2025,Cao2025,Cerezo2022,Mohseni2022,Rosati2023,Bilkis2021,Rosati2022}.

%Altogether, our results identify a quantum-native formulation of SA in which the nonlinearity arises from interference of overlaps and the training objective is a direct measurement, positioning QSA as a modular primitive for larger quantum machine learning pipelines~\cite{Puig2025,Cao2025,Cerezo2022,Mohseni2022,Rosati2023,Bilkis2021,Rosati2022}, for both classical and quantum sequences, where data similarity is intrinsically an overlap of states.

\ssection{Classical self-attention} %{\color{red} and next-token distribution}}
In a standard language prediction task, words $\bw_1,\ldots,\bw_{T+1}$ of a length-$(T+1)$ sequence are embedded into {\color{red} feature-space} tokens $\tilde\bx_i\in\mathbb{R}^d$ as $\tilde\bx_i = E \bw_i + \bc_i$, where the rectangular $E\in \R^{d\times D}$ is a linear embedding layer mapping one-hot-encoded vocabulary words $\bw_i\in\cV$, $\cV:= \{\bw^{(\ell)}=(0_1,\cdots,0_{\ell-1},1_\ell,0_{\ell+1},\cdots,0_D)\}_{\ell=1}^D$, into a feature space of much smaller dimension $d\ll D$, and $\bc_i$ are shifts encoding positional information.  

For each training sentence the SA output is a weighted sum of tokens up to $j$, with affinity weights given by the tokens' inner products: $\tilde \bz_j = \sum_{i=1}^j \mathrm{softmax}\!\left(\frac{\bq_j \cdot \bk_i }{\sqrt d}\right)\bv_i$, where $\bv_i,\bk_i,\bq_i$ are the value, key and query vectors, obtained by distinct linear layers applied to $\tilde \bx_i$, and the softmax acts with respect to $i$ at fixed $j$~\cite{Vaswani}. 
Classical transformers then apply a residual connection, i.e., a shift by $\bx_i$, and a feed-forward network with linear layers and non-linear activation, producing a token $\bz'_i$; finally an anti-embedding layer, i.e., a linear transformation plus a softmax, produces a probability vector whose components $\ell=1,\cdots,D$ are the conditional probabilities ${\color{red}p_{j+1}^{(\ell)}}:=P(\bw_{j+1}=\bw^{(\ell)}\,|\,\bw_1,\ldots,\bw_j)$ of predicting the $\ell$-th word of the vocabulary at the $(j+1)$-th step. During training, the true next word in the sequence is known, {\color{red} i.e. $\ell_{j+1}$,} hence the performance is evaluated via the {\color{red} Shannon cross-entropy loss: $\cL_1 := {\rm CE}_{1}(u_T||p) \equiv -\frac{1}{T}\sum_{j=1}^{T}\log (p_{j+1}^{(\ell_{j+1})})$, where $u_T$ is the uniform distribution on $T$ points and we employ base-$2$ logarithms. }

{\color{red}Implementing analogous transformations on a quantum computer faces non-linearity issues from three sources: (i) the non-linear combination of tokens giving the SA affinity weights; (ii) the softmax normalization of tokens within SA and at the output; (iii) post-processing feedforward nonlinearities. Sources (i) and (ii) are intrinsic to SA---the former combines tokens by affinity, the latter exponentially enhances or suppresses attention weights---whereas (iii) is a generic expressivity device of neural architectures.

Our construction realizes both intrinsic nonlinearities: the affinity weights exactly, and the softmax as a degree-$k$ monomial (mono-QSA) or polynomial (poly-QSA) kernel $f_k(\cdot) = \sum_{p=0}^k c_p (\cdot)^k$ with controlled truncation error. Only the feedforward block post-processing the attention output remains outside it. The two are structurally different: the softmax nonlinearity acts on the \textit{overlaps} $\bq_j\cdot\bk_j$, which the interference readout produces natively, whereas the feedforward block applies a \textit{pointwise} map to the amplitudes of $\tilde\bz_j$; extending the construction that way requires a different mechanism, e.g., a specific expressivity-increasing quantum circuit, left to future work. %— for instance treating QSA as a subroutine within a larger circuit, or invoking nonlinear-transform primitives such as QSVT —

We thus redefine the SA output as $ \bz_j =\sum_{i=1}^j f_k(\bq_j\cdot\bk_i)\, \bv_i$, up to normalization. The post-SA steps then aggregate into a linear layer and a shift, mapping $\bz_j$ to a prediction vector $\hat\bw_j \in\R^D$ as $\hat\bw_j = F(\bz_j + \tilde\bx_j)$, with $F\in\R^{D\times d}$ a rectangular anti-embedding matrix. By the one-hot encoding, the probability of predicting the $\ell$-th word $\bw^{(\ell)}$ at the next step is $p_{j+1}^{(\ell)} = \left(\frac{\bw^{(\ell)}\cdot\hat\bw_{j}}{||\hat\bw_j ||_2}\right)^2 = \frac{(\hat\bw_j)_\ell^2}{\sum_{\ell=1}^D (\hat\bw_j)_\ell^2}$, which is manifestly positive and normalized across words, i.e., $\sum_{\ell=1}^D p_{j+1}^{(\ell)}=1$.
Expressing the $\ell$-th word via its embedded token, $\bw^{(\ell)}=E^{-1}(\tilde\bx^{(\ell)}-\bc^{(\ell)})$, gives $(\hat\bw_j)_\ell = (\tilde\bx^{(\ell)}-\bc^{(\ell)})^T E^{-1 T} F(\bz_j + \tilde \bx_j)$, a sum of two terms: the $\bz_j$-dependent one is an inner product evaluated directly in feature space from the SA output; the $\tilde \bx_j$-dependent one is independent of $\bz_j$ and amounts to an overall shift, absorbable into $\bc_j$. Hence, absent feedforward non-linearities, the residual connection can be set to zero without loss of generality; with shifted tokens $\bx^{(\ell)} = \tilde\bx^{(\ell)}-\bc^{(\ell)}$ and $G=(E^{-1})^{T} F$, the prediction probability reads
\begin{equation}\label{eq:prediction_probability_classical_final}
    p_{j+1}^{(\ell)} = \frac{(\bx^{(\ell) T} G \bz_j)^2}{\sum_{\ell'=1}^D (\bw^{(\ell')T}F \bz_j)^2} = \left(\frac{\bx^{(\ell) T} G \bz_j}{||F\bz_j||_2}\right)^2,
\end{equation}
having used that one-hot words form an orthonormal basis, $\sum_\ell \bw^{(\ell)} \bw^{(\ell)T} = \1_D$. For isometric $F$, as we choose for our QSA, the denominator reduces to $||\bz_j||_2^2$.
}

\ssection{Quantum self-attention}
Our QSA rests on a key observation: up to normalization, {\color{red} the probability \eqref{eq:prediction_probability_classical_final} is polynomial in the input tokens and their inner products. Hence, amplitude-encoding~\cite{Schuld2018h} the normalized tokens $\ket{\bxi_j} := U_j \ket 0 \equiv\ket{\bx_j}/\sqrt{a_j}$, with $a_j = \braket{\bx_j}{\bx_j}$, into quantum state vectors, we construct a circuit that efficiently calculates quantities akin to~\eqref{eq:prediction_probability_classical_final} in superposition, as described below, in Fig.~\ref{fig:circuit} and detailed in the supplemental material (SM). We write $\bx_{j+1}=\bx^{(\ell_{j+1})}$ for the unnormalized correct next token, and likewise $\bxi_{j+1}$, $a_{j+1}$ for its normalized version and norm.

\begin{figure}
    \centering
    \includegraphics[width=0.45\textwidth,trim={.1cm 3cm .5cm 7.8cm},clip]{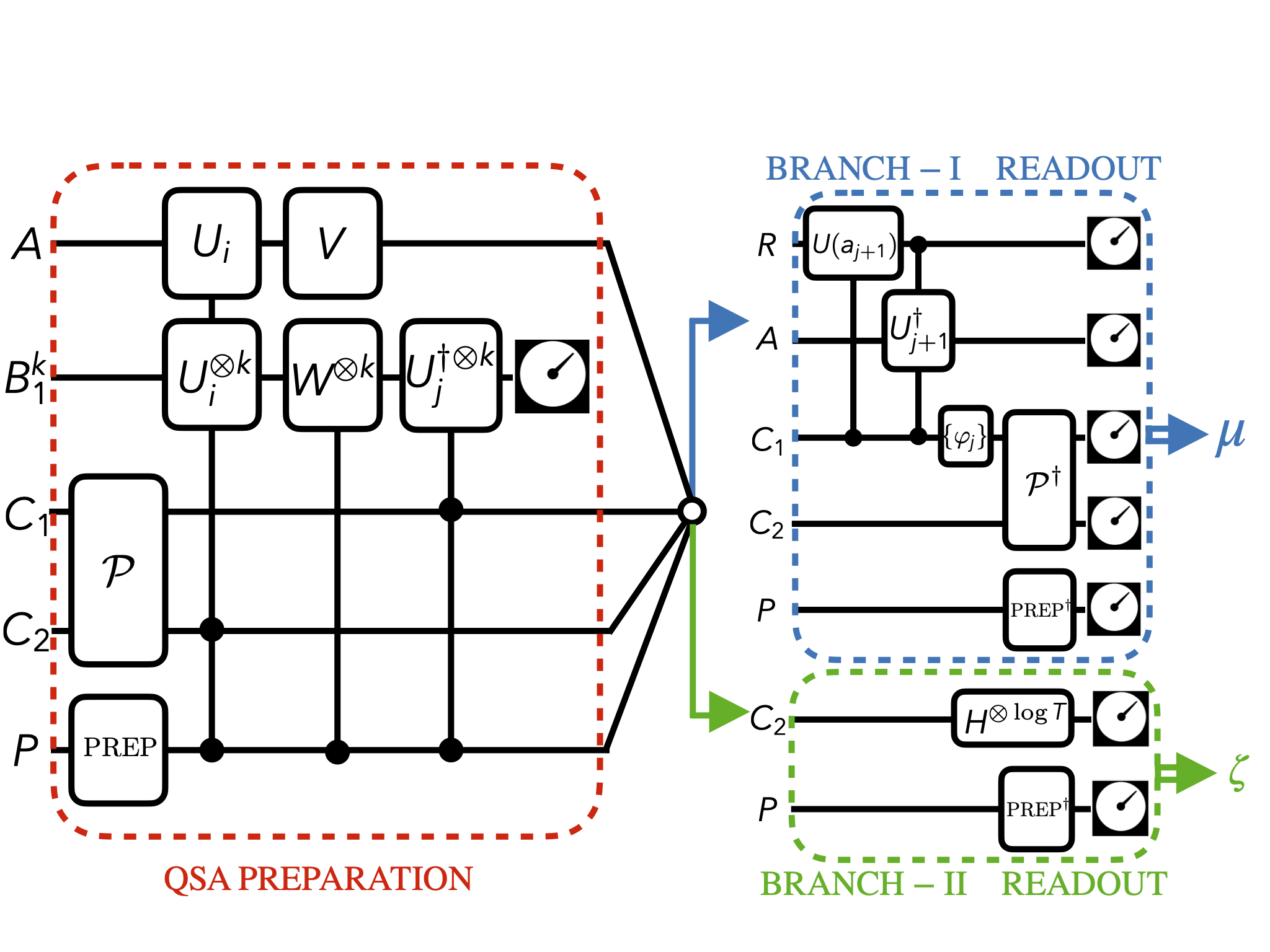}
    \caption{Schematic depiction of our QSA preparation and the two readout circuits, as described in the text. }
    \label{fig:circuit}
\end{figure}

\begin{figure*}
    \centering
    \includegraphics[width=0.45\linewidth]{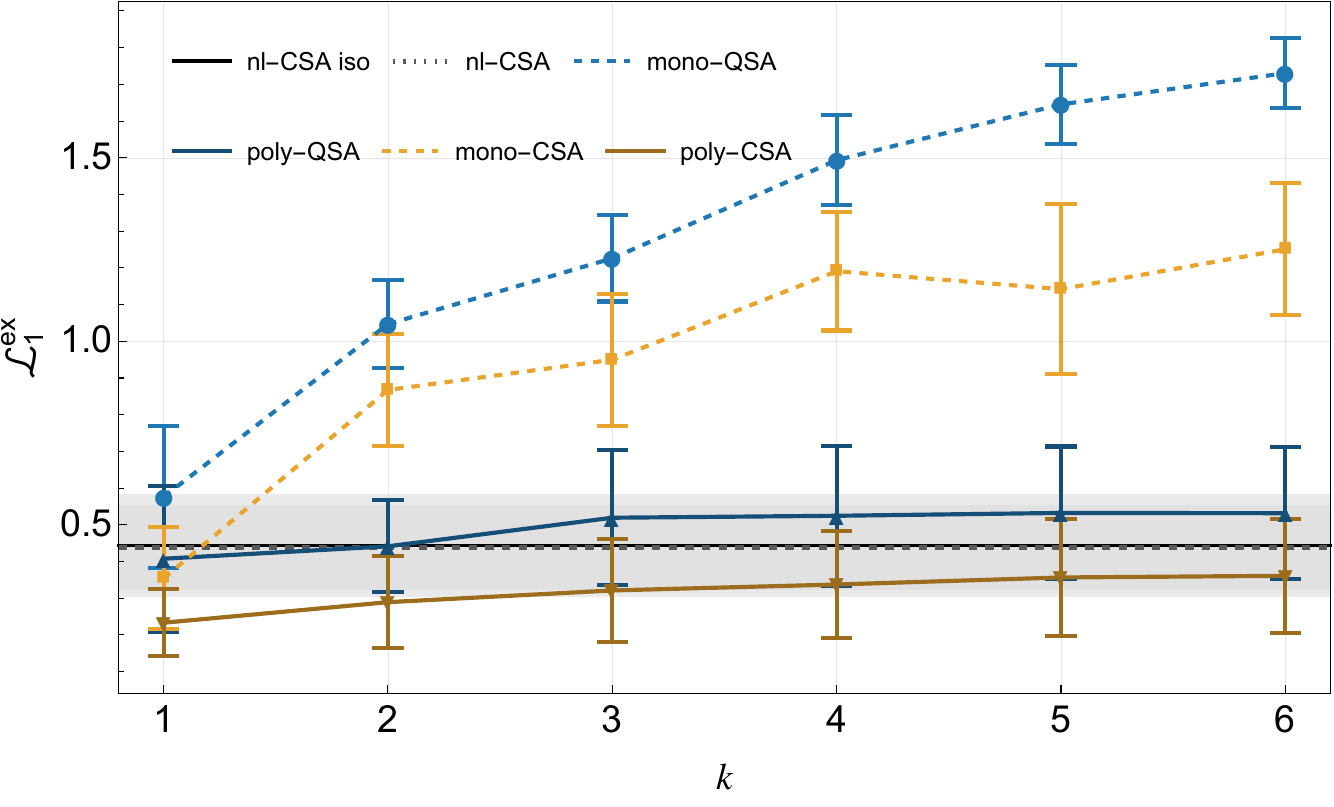}
     \includegraphics[width=0.45\linewidth]{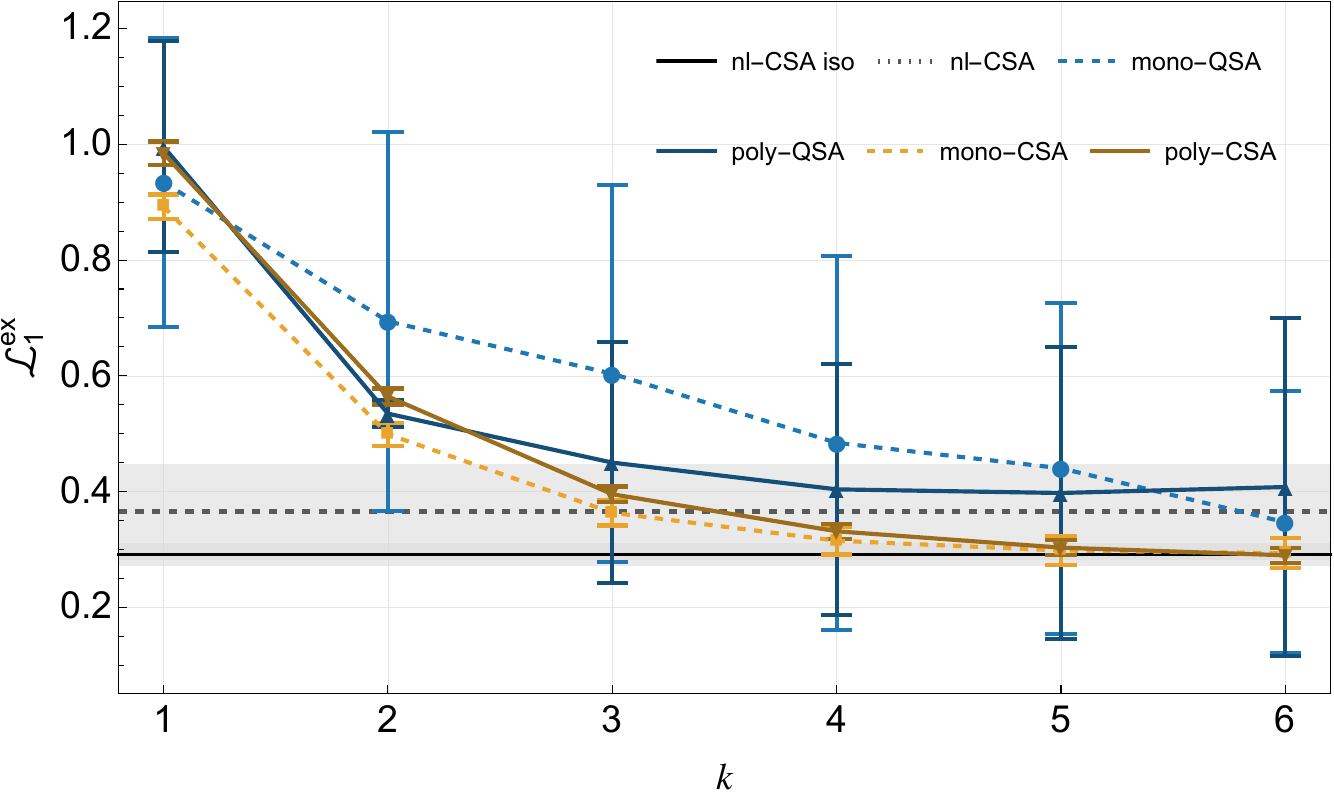}
    \caption{Plot of $\cL_1^{\rm ex}$ after training for mono/poly-QSA, -CSA, and nl-CSA models, vs the degree $k$ ( $8$-run averages). Number of trainable parameters: 1032 (QSA), 1024 (CSA). (left) classical sequences; (right) quantum sequences.}
    \label{fig:training_curves}
\end{figure*}

The QSA circuit runs on five registers: (i) $k+1$ data registers $A$, $B_1^k$ of $\log d$ qubits each, holding respectively the values and $k$ copies of the query-key overlaps; (ii) timestep- and token-control registers $C_{1,2}$ of $\log T$ qubits each; (iii) for poly-QSA, a degree-control register $P$ of $k$ qubits \footnote{These can be brought down to $\log k$ by binary encoding, at the cost of comparator operations for the controlled gates; for modest $k$ we favour the qubit over the gate overhead.}. The control registers are initially prepared, via ${\rm PREP}_P$ and $\cP_{C_{1,2}}$, in the superposition $\sum_{p=0}^{k}\sqrt{\frac{c_p}{\lambda}}\;\ket{1^p0^{k-p}}_P \otimes\frac{1}{\sqrt{\cN}}\sum_{j=1}^{T}\sum_{i=1}^j\ket{j}_{C_1}\ket{i}_{C_2}$, with normalizations $\lambda = \sum_{p=0}^k c_p$, $\cN = \binom{T+1}{2}$ and $\{\ket j\}_{j=1}^{T}$ the computational-basis binary representation of $j$. For poly-QSA, operations on $B$ are controlled by $P$, implementing a labelled superposition of degree-$p$ monomials for $p\leq k$.

The core attention mechanism follows in three operations: (i) amplitude-encoding of $\ket{\bxi_i}^{\otimes (k+1)}$ on $A B_1^k$ in superposition for all $i\leq j$, controlled by $C_2$; (ii) variational gates $V_A \otimes (W^{\otimes k})_{B_1^k}$ on the data registers, analogous to the classical linear layers generating values, keys and queries; (iii) uncomputation of the query-key tokens, yielding the attention weights, via a projection $(\ket{\bx_j}\bra{\bx_j})^{\otimes k}$ on $B_1^k$ controlled by $C_1$. Conditional on the all-zero outcome on $B_1^k$, the resulting state superposes the value vectors $V\ket{\bxi_j}$ with weighted powers of the query-key overlaps, i.e., the affinity scores: $\sum_{p,j,i}\sqrt{\frac{c_p}{\lambda \cN}}(\bra{\bxi_{j}}W\ket{\bxi_i})^p\, V\ket{\bxi_i} \otimes\ket{1^p 0^{k-p}, j,i}_{P C_{1,2}}$.
%and on the next token $\bra{\bx_{j+1}}$ for register $A$, realized by uncomputation of the amplitude-encoding gates controlled by the sequence register $C_1$. 

Two expectation values are then extracted via suitable readout circuits. \textit{Branch-I:} uncompute the target token by measuring $\dketbra{\bx_{j+1}}$ on $A$ controlled by $C_1$; as $\ket{\bx}$ is unnormalized, we implement it via projection on $\dketbra{\bxi_{j+1}}$ controlled by an extra one-qubit register $R$ in the state $U(a_{j+1})\ket{0}=\sqrt{a_{j+1}}\ket{0} + \sqrt{1-a_{j+1}}\ket{1}$. Apply trainable phases $\{\varphi_j\}_{j=1}^T$ to $C_1$ and, finally, uncompute the control registers' preparations, yielding the all-zero expectation $\mu := \left|\sum_{j=1}^T e^{\ii \varphi_j} \braket{\bx_{j+1}}{\bz_j} \right|^2/(\lambda\, \cN)^2$,
where $\ket{\bz_j} := \sum_{p,i} (\bra{\bx_{j}}W\ket{\bx_j})^p\, V\ket{\bx_j}$ are the unnormalized prediction states. 
\textit{Branch-II:} discard $C_1 A$, apply ${\rm PREP}_P^\dag$ and a uniform-weight uncomputation of $C_2$, i.e., a projection on $(H \ket{0})^{\otimes \log T}$, yielding the all-zero expectation $ \zeta := \frac{\sum_{j=1}^T \braket{\bz_j}{\bz_j}}{\lambda^2 \cN T}$.

\ssection{\color{red}  R\'enyi loss analysis} The QSA training loss is defined as $\cL_{\rm qsa} := -\log\frac{(T+1) \mu}{2T \zeta}$, whose properties we now investigate, deferring proofs to the SM. First, it admits an interpretation in terms of R\'enyi entropic quantities:
\begin{theorem}\label{thm:loss_entropic}
    If $G$ is unitary and the trainable phases are aligned to the overlaps of correct-tokens and prediction, i.e., $\varphi_j=-{\rm arg}\braket{\bx_{j+1}}{\bz_j}$ for all $j=1,\cdots,T$, then
    \begin{equation}\label{eq:loss_entropic}
        \cL_{\rm qsa} = {\rm CE}_{1/2}(q||p) + D_{1/2}(u_T||q),
    \end{equation}
    where $D_\alpha(q||p)=\frac{1}{\alpha-1}\log\sum_j q_j^{\alpha}p_j^{1-\alpha}$,  ${\rm CE}_\alpha(q||p) = D_\alpha(q||p) + H_\alpha(q)$ and $H_\alpha(q)=\frac{1}{1-\alpha}\log\sum_j q_j^{\alpha}$ are the R\'enyi-$\alpha$ divergence, cross-entropy (see~\cite{Thierrin2022}), and entropy, $q_j = ||\bz_j||_2^2/\sum_{j'}||\bz_{j'}||_2^2$ is the distribution of prediction-norms across time-steps, and  $p_j=p_{j+1}^{(\ell_{j+1})}$ is the correct-next-token probability of \eqref{eq:prediction_probability_classical_final} (which is not normalized across $j$).  
\end{theorem}
Let us comment on the assumptions of Thm.~\ref{thm:loss_entropic}. Unitary $G$ is a design choice constraining $E^{-1}$ and $F$ to be isometries onto the same subspace of $\R^D$, i.e., ${\rm ran}\, E^{-1} = {\rm ran}\, F$ (e.g., $E^{-1}=F$, which we do not assume). It merely simplifies the circuit, merging $G$ into $V$, and can be dropped at the cost of ancillary qubits, embedding a non-unitary $G$ into a larger circuit and a non-isometric $F$ in the branch-2 readout to obtain $F\ket{\bz_j}$. Secondly, at fixed values of the other parameters the phase-alignment condition is the unique minimum of $\cL_{\rm qsa}$ (see SM), so the regime of Thm.~\ref{thm:loss_entropic} is accessible during training.

Then, Eq.~\eqref{eq:loss_entropic} splits the loss into two meaningful terms: $ D_{1/2}(u_T||q)$ penalizes prediction-norm distributions far from uniform, since they give more weight to specific time-steps; ${\rm CE}_{1/2}(q||p)$ penalizes models whose correct-next-token probability is small relative to the prediction norms. Both are non-negative, so setting them to zero minimizes $\cL_{\rm qsa}$. In particular $ D_{1/2}(u_T||q)=0$ iff $q=u_T$, i.e., equal weight on each time-step as in a standard generative task, and the loss reduces to ${\rm CE}_{1/2}(u_T||p)$, akin to the classical $\cL_1$. Indeed the two losses are closely related, and minimizing $\cL_{\rm qsa}$ implies minimizing $\cL_1$:
\begin{corollary}
    Under the assumptions of Thm.~\ref{thm:loss_entropic} , it holds $\cL_{\rm qsa} \leq \cL_1 + D_1(u_T||q)$, which is saturated at $\cL_{\rm qsa}=0 \Rightarrow \cL_1=0$, the converse being false.
\end{corollary}
Importantly, neither loss can attain zero in general, since the embedding does not intrinsically normalize tokens $\ket{\bx_{j+1}}$: indeed $\cL_{\rm qsa}\geq -\log a_A$ and $\cL_1\geq -\log a_G$, with $a_A = \tfrac1T \sum_j a_{j+1}$, $a_G = (\prod_j a_{j+1})^{1/T}$ the arithmetic and geometric means of the correct-next-token norms. Defining the excess losses $\cL_{\rm qsa}^{\rm ex} := \cL_{\rm qsa} + \log a_A$,  $\cL_1^{\rm ex} :=\cL_1 + \log a_G$ and the token-norms' distribution $\tilde a_j = a_{j+1}/(T a_A)$ yields zero-floor losses akin to $\cL_{{\rm qsa},1}$:
\begin{corollary}
    Under the assumptions of Thm.~\ref{thm:loss_entropic} , it holds $\cL_{\rm qsa}^{\rm ex} \leq \cL_1^{\rm ex} + D_1(u_T||q) + D_1(u_T||\tilde a)$ and $\cL_{\rm qsa}^{\rm ex}=0 \Rightarrow \cL_1^{\rm ex}=0$, the converse being false. The zero-point saturates the inequality only in the balanced case $\tilde a = u_T$. 
\end{corollary}
We note that the token norms obey the global constraint $\sum_{\ell=1}^D a_\ell = \Tr[E^\dagger E ]$, equal to $d$ for isometric $E$, so allocating norms to words is non-trivial. We therefore train on $\cL_{\rm qsa}$, whose optimization depends on the embedding norms, but report the final performance via excess losses, whose zero-floor accounts for the fixed norms.  
 }

\ssection{Generative modelling of classical and quantum data sequences}
We parameterize $V$ and $W$ with {\color{red} $L$ layers of two-body interactions (as in~\cite[Fig.~2]{LaRose2019}), of gate complexity $O(L \log d)$}, and test QSA on classical and quantum sequence-prediction tasks, training on sequences {\color{red} of length up to $T=32$ and embedding dimension up to $d=32$}. As baselines we train, at the same embedding dimension, {\color{red} a standard non-linear CSA (nl-CSA) with softmax attention and unconstrained value/anti-embedding, as well as one with isometric value/anti-embedding (nl-CSA iso), both on the $\cL_1$ loss, and a classical twin (mono/poly-CSA) of our QSA on $\cL_{\rm qsa}$. Classical data are sampled from the Penn Treebank dataset; quantum data from a random $\log D$-qubit transverse-field Ising Hamiltonian $H=-J\sum_{i=1}^{\log D-1} Z_i Z_{i+1}-h\sum_{i=1}^{\log D} X_i$
%$H=\sum_{i=1}^D X_i+\sum_{i<j}J_{ij}Z_iZ_j$
with $J, h\in[0.2,2]$ and $D=16$. From Haar-random initial states $\ket{\psi_1}$ we generate $\ket{\psi_j}=e^{-\ii H(j-1)\Delta t}\ket{\psi_1}$, $j=2,\ldots,T+1$, $\Delta t=0.35$, identifying them with tokens $\ket{\bx_j}$, without embedding.

Fig.~\ref{fig:training_curves}a,b plot the final training-set loss of all models, evaluated with $\cL_1^{\rm ex}$ to allow a fair comparison with nl-CSA, against the degree $k$. For classical sequences polynomial models quickly reach loss values comparable to the non-linear classical ones, while monomial models remain considerably higher; for quantum sequences both monomial and polynomial models approach the non-linear performance as $k$ grows. The classical twins follow QSA's behaviour albeit with a favourable gap.

These results reflect the nature of the data and of the kernel. % rather than QSA-specific features.
For quantum sequences the overlaps are state fidelities, so raising the kernel degree acts as an increasingly selective filter on states aligned with the correct one, for both monomial and polynomial kernels. Instead, for classical data the token overlaps are set by the learned embedding and lack such an interpretation: useful contextual information may be spread over many moderate overlaps, progressively suppressed by the monomial kernel, whereas the polynomial kernel retains lower-degree terms and so preserves affinity with the wider context.
}

\ssection{\color{red} Training and sampling complexity}
{\color{red} As shown in the SM, our QSA has an end-to-end training complexity $O(\mu^{-1} k^2 T d \log d )$, dominated by amplitude-encoding \textit{without QRAM}, i.e., $O(d)$ per token $\ket{\bxi}$~\cite{Schuld2018h}. The $\mu$-dependence accounts for the shots required for expectations and derivatives via parameter-shift rule~\cite{Schuld2019,Mari2021}, and the ansatz cost assumes $O(1)$ layers with $O(\log d)$ parametrized gates. The $O(TdD)$ embedding cost is not counted, being a baseline classical cost common to all studied models.
By contrast, standard softmax self-attention (nl-CSA) scales as $O(T^2d)$~\cite{Vaswani}, quadratically worse in the sequence length. A fairer comparison is the classical twin mono/poly-CSA, which implements QSA's kernel at cost $O(T d m_k)$ with $m_k = \binom{d+k-1}{k}\sim d^k$: it is cheaper than nl-CSA for long-sequences $T\gg m_k$, and already well-approximates softmax at small $k$.

Comparing these complexities, a quantum advantage arises when $\mu \geq \mu_{\rm adv}(k,d):= O(m_k^{-1} k^2 \log d)$. Fig.~\ref{fig:advantage} shows that $\mu$ for our trained models saturates in $T$ and closes the gap to $\mu_{\rm adv}$ as $d$ grows; extrapolating to the asymptotic regime predicts an advantage at $d\gtrsim 119$ for $k=7$, confirming that the long-sequence advantage regime is accessible in practice.

%Sampling from the trained model might proivde an advantage with respect to classical techniques, but only provided that the effective success probability increases significantly with respect to its random-token estimate (see SM).}

\begin{figure}
    \centering
    \includegraphics[width=0.45\textwidth,trim={.1cm .5cm .1cm 2cm},clip]{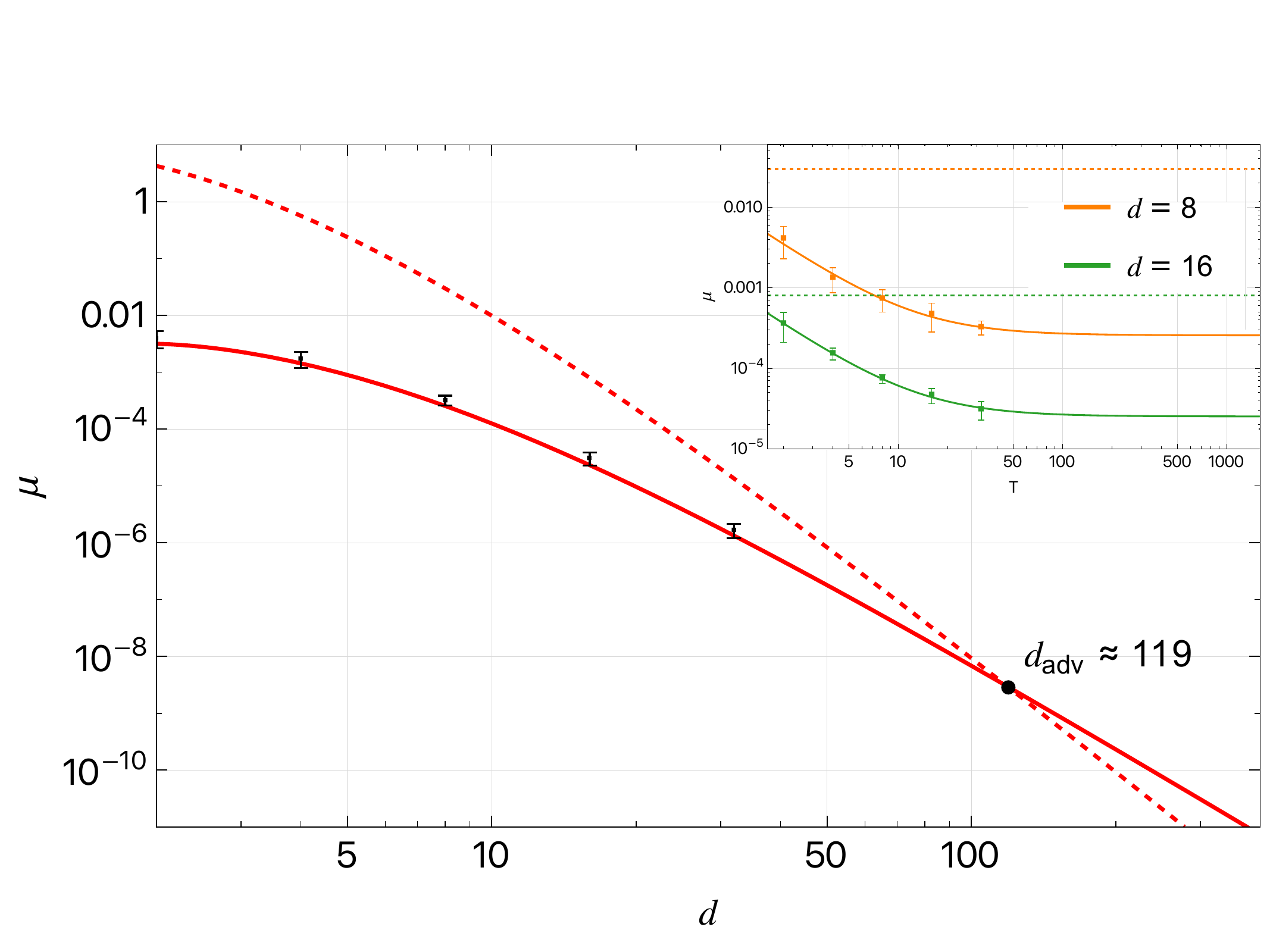}
    \caption{Plots of trained $\mu$ (points, $10$-run average) and the advantage threshold $\mu_{\rm adv}(k,d)$ (dashed) vs d for $T=32$, $k=7$. The solid line originates from a power-law fit of the ratio $\mu_{\rm adv}/\mu$, extrapolating the large-$d$ limit. Inset: same quantities vs $T$, proving saturation at large $T$. }
    \label{fig:advantage}
\end{figure}

\ssection{Discussion and conclusions} 
We introduced a variational polynomial-kernel quantum self-attention (QSA) in which attention is implemented quantum-natively via quantum states' inner products, and the loss is built from two observables, bypassing the decoding bottleneck of previous works. Numerical tests show performance comparable to classical models on both classical and many-body quantum sequences, and support a training-complexity advantage in the long-sequence regime. We therefore expect QSA to become an enabling primitive for quantum transformer architectures and a practical tool for attention-driven modeling of many-body quantum systems.
}

\ssection{Acknowledgements}
M.R. acknowledges support by MUR (project FIS-2023-03472, “SuNRISE” CUP F53C24001760001), CINECA HPC ISCRA C projects "QuSALa" and "ManQO-GPT", and M. Barbieri. The data generated for this research are available at \href{https://github.com/Alessio-Pecilli/variational_quantum_transformer_superposition/tree/PennyLaneG}{Github}.%.R. acknowledges support from the project PNRR - Id MSCA 0000011-SQUID - CUP F83C22002390007 (Young Researchers) - Finanziato dall’Unione europea - NextGenerationEU.

\bibliography{library,library_addition}

\begin{thebibliography}{22}%
\makeatletter
\providecommand \@ifxundefined [1]{%
 \@ifx{#1\undefined}
}%
\providecommand \@ifnum [1]{%
 \ifnum #1\expandafter \@firstoftwo
 \else \expandafter \@secondoftwo
 \fi
}%
\providecommand \@ifx [1]{%
 \ifx #1\expandafter \@firstoftwo
 \else \expandafter \@secondoftwo
 \fi
}%
\providecommand \natexlab [1]{#1}%
\providecommand \enquote  [1]{``#1''}%
\providecommand \bibnamefont  [1]{#1}%
\providecommand \bibfnamefont [1]{#1}%
\providecommand \citenamefont [1]{#1}%
\providecommand \href@noop [0]{\@secondoftwo}%
\providecommand \href [0]{\begingroup \@sanitize@url \@href}%
\providecommand \@href[1]{\@@startlink{#1}\@@href}%
\providecommand \@@href[1]{\endgroup#1\@@endlink}%
\providecommand \@sanitize@url [0]{\catcode `\\12\catcode `\$12\catcode `\&12\catcode `\#12\catcode `\^12\catcode `\_12\catcode `\%12\relax}%
\providecommand \@@startlink[1]{}%
\providecommand \@@endlink[0]{}%
\providecommand \url  [0]{\begingroup\@sanitize@url \@url }%
\providecommand \@url [1]{\endgroup\@href {#1}{\urlprefix }}%
\providecommand \urlprefix  [0]{URL }%
\providecommand \Eprint [0]{\href }%
\providecommand \doibase [0]{https://doi.org/}%
\providecommand \selectlanguage [0]{\@gobble}%
\providecommand \bibinfo  [0]{\@secondoftwo}%
\providecommand \bibfield  [0]{\@secondoftwo}%
\providecommand \translation [1]{[#1]}%
\providecommand \BibitemOpen [0]{}%
\providecommand \bibitemStop [0]{}%
\providecommand \bibitemNoStop [0]{.\EOS\space}%
\providecommand \EOS [0]{\spacefactor3000\relax}%
\providecommand \BibitemShut  [1]{\csname bibitem#1\endcsname}%
\let\auto@bib@innerbib\@empty
%</preamble>
\bibitem [{\citenamefont {Calvino}\ \emph {et~al.}(2025)\citenamefont {Calvino}, \citenamefont {Reijerink},\ and\ \citenamefont {Samek}}]{Calvino2025}%
  \BibitemOpen
  \bibfield  {author} {\bibinfo {author} {\bibfnamefont {F.}~\bibnamefont {Calvino}}, \bibinfo {author} {\bibfnamefont {J.}~\bibnamefont {Reijerink}},\ and\ \bibinfo {author} {\bibfnamefont {L.}~\bibnamefont {Samek}},\ }\bibfield  {title} {\bibinfo {title} {{The effects of generative AI on productivity, innovation and entrepreneurship}},\ }\bibfield  {journal} {\bibinfo  {journal} {OECD Artif. Intell. Pap.}\ }\bibinfo {series} {OECD Artificial Intelligence Papers},\ \href {https://doi.org/10.1787/b21df222-en} {10.1787/b21df222-en} (\bibinfo {year} {2025})\BibitemShut {NoStop}%
\bibitem [{\citenamefont {Vaswani}\ \emph {et~al.}(2017)\citenamefont {Vaswani}, \citenamefont {Shazeer}, \citenamefont {Parmar}, \citenamefont {Uszkoreit}, \citenamefont {Jones}, \citenamefont {Gomez}, \citenamefont {Kaiser},\ and\ \citenamefont {Polosukhin}}]{Vaswani}%
  \BibitemOpen
  \bibfield  {author} {\bibinfo {author} {\bibfnamefont {A.}~\bibnamefont {Vaswani}}, \bibinfo {author} {\bibfnamefont {N.}~\bibnamefont {Shazeer}}, \bibinfo {author} {\bibfnamefont {N.}~\bibnamefont {Parmar}}, \bibinfo {author} {\bibfnamefont {J.}~\bibnamefont {Uszkoreit}}, \bibinfo {author} {\bibfnamefont {L.}~\bibnamefont {Jones}}, \bibinfo {author} {\bibfnamefont {A.~N.}\ \bibnamefont {Gomez}}, \bibinfo {author} {\bibfnamefont {L.}~\bibnamefont {Kaiser}},\ and\ \bibinfo {author} {\bibfnamefont {I.}~\bibnamefont {Polosukhin}},\ }\bibfield  {title} {\bibinfo {title} {{Attention Is All You Need}},\ }\href {https://arxiv.org/abs/1706.03762 http://arxiv.org/abs/1706.03762} {\  (\bibinfo {year} {2017})},\ \Eprint {https://arxiv.org/abs/1706.03762} {arXiv:1706.03762} \BibitemShut {NoStop}%
\bibitem [{\citenamefont {Cherrat}\ \emph {et~al.}(2022)\citenamefont {Cherrat}, \citenamefont {Kerenidis}, \citenamefont {Mathur}, \citenamefont {Landman}, \citenamefont {Strahm},\ and\ \citenamefont {Li}}]{Cherrat2022}%
  \BibitemOpen
  \bibfield  {author} {\bibinfo {author} {\bibfnamefont {E.~A.}\ \bibnamefont {Cherrat}}, \bibinfo {author} {\bibfnamefont {I.}~\bibnamefont {Kerenidis}}, \bibinfo {author} {\bibfnamefont {N.}~\bibnamefont {Mathur}}, \bibinfo {author} {\bibfnamefont {J.}~\bibnamefont {Landman}}, \bibinfo {author} {\bibfnamefont {M.}~\bibnamefont {Strahm}},\ and\ \bibinfo {author} {\bibfnamefont {Y.~Y.}\ \bibnamefont {Li}},\ }\bibfield  {title} {\bibinfo {title} {{Quantum Vision Transformers}},\ }\href {https://doi.org/10.22331/q-2024-02-22-1265} {\bibfield  {journal} {\bibinfo  {journal} {Quantum}\ }\textbf {\bibinfo {volume} {8}},\ \bibinfo {pages} {1265} (\bibinfo {year} {2022})},\ \Eprint {https://arxiv.org/abs/2209.08167} {arXiv:2209.08167} \BibitemShut {NoStop}%
\bibitem [{\citenamefont {Gao}\ \emph {et~al.}(2023)\citenamefont {Gao}, \citenamefont {Song}, \citenamefont {Yang},\ and\ \citenamefont {Zhang}}]{Gao2023}%
  \BibitemOpen
  \bibfield  {author} {\bibinfo {author} {\bibfnamefont {Y.}~\bibnamefont {Gao}}, \bibinfo {author} {\bibfnamefont {Z.}~\bibnamefont {Song}}, \bibinfo {author} {\bibfnamefont {X.}~\bibnamefont {Yang}},\ and\ \bibinfo {author} {\bibfnamefont {R.}~\bibnamefont {Zhang}},\ }\bibfield  {title} {\bibinfo {title} {{Fast Quantum Algorithm for Attention Computation}},\ }\href {http://arxiv.org/abs/2307.08045} {\  (\bibinfo {year} {2023})},\ \Eprint {https://arxiv.org/abs/2307.08045} {arXiv:2307.08045} \BibitemShut {NoStop}%
\bibitem [{\citenamefont {Zhao}\ \emph {et~al.}(2024{\natexlab{a}})\citenamefont {Zhao}, \citenamefont {Shi},\ and\ \citenamefont {Li}}]{Zhao2024}%
  \BibitemOpen
  \bibfield  {author} {\bibinfo {author} {\bibfnamefont {R.-X.}\ \bibnamefont {Zhao}}, \bibinfo {author} {\bibfnamefont {J.}~\bibnamefont {Shi}},\ and\ \bibinfo {author} {\bibfnamefont {X.}~\bibnamefont {Li}},\ }\bibfield  {title} {\bibinfo {title} {{GQHAN: A Grover-inspired Quantum Hard Attention Network}},\ }\href {http://arxiv.org/abs/2401.14089} {\  (\bibinfo {year} {2024}{\natexlab{a}})},\ \Eprint {https://arxiv.org/abs/2401.14089} {arXiv:2401.14089} \BibitemShut {NoStop}%
\bibitem [{\citenamefont {Zhao}\ \emph {et~al.}(2024{\natexlab{b}})\citenamefont {Zhao}, \citenamefont {Shi},\ and\ \citenamefont {Li}}]{Zhao2024a}%
  \BibitemOpen
  \bibfield  {author} {\bibinfo {author} {\bibfnamefont {R.-X.}\ \bibnamefont {Zhao}}, \bibinfo {author} {\bibfnamefont {J.}~\bibnamefont {Shi}},\ and\ \bibinfo {author} {\bibfnamefont {X.}~\bibnamefont {Li}},\ }\bibfield  {title} {\bibinfo {title} {{QKSAN: A Quantum Kernel Self-Attention Network}},\ }\href {https://doi.org/10.1109/TPAMI.2024.3434974} {\bibfield  {journal} {\bibinfo  {journal} {IEEE Trans. Pattern Anal. Mach. Intell.}\ }\textbf {\bibinfo {volume} {46}},\ \bibinfo {pages} {10184} (\bibinfo {year} {2024}{\natexlab{b}})}\BibitemShut {NoStop}%
\bibitem [{\citenamefont {Liao}\ and\ \citenamefont {Ferrie}(2024)}]{Liao2024}%
  \BibitemOpen
  \bibfield  {author} {\bibinfo {author} {\bibfnamefont {Y.}~\bibnamefont {Liao}}\ and\ \bibinfo {author} {\bibfnamefont {C.}~\bibnamefont {Ferrie}},\ }\bibfield  {title} {\bibinfo {title} {{GPT on a Quantum Computer}},\ }\href {http://arxiv.org/abs/2403.09418} {\  (\bibinfo {year} {2024})},\ \Eprint {https://arxiv.org/abs/2403.09418} {arXiv:2403.09418} \BibitemShut {NoStop}%
\bibitem [{\citenamefont {Guo}\ \emph {et~al.}(2024)\citenamefont {Guo}, \citenamefont {Yu}, \citenamefont {Choi}, \citenamefont {Agrawal}, \citenamefont {Nakaji}, \citenamefont {Aspuru-Guzik},\ and\ \citenamefont {Rebentrost}}]{Guo2024}%
  \BibitemOpen
  \bibfield  {author} {\bibinfo {author} {\bibfnamefont {N.}~\bibnamefont {Guo}}, \bibinfo {author} {\bibfnamefont {Z.}~\bibnamefont {Yu}}, \bibinfo {author} {\bibfnamefont {M.}~\bibnamefont {Choi}}, \bibinfo {author} {\bibfnamefont {A.}~\bibnamefont {Agrawal}}, \bibinfo {author} {\bibfnamefont {K.}~\bibnamefont {Nakaji}}, \bibinfo {author} {\bibfnamefont {A.}~\bibnamefont {Aspuru-Guzik}},\ and\ \bibinfo {author} {\bibfnamefont {P.}~\bibnamefont {Rebentrost}},\ }\bibfield  {title} {\bibinfo {title} {{Quantum linear algebra is all you need for Transformer architectures}},\ }\href {http://arxiv.org/abs/2402.16714} {\  (\bibinfo {year} {2024})},\ \Eprint {https://arxiv.org/abs/2402.16714} {arXiv:2402.16714} \BibitemShut {NoStop}%
\bibitem [{\citenamefont {Khatri}\ \emph {et~al.}(2024)\citenamefont {Khatri}, \citenamefont {Matos}, \citenamefont {Coopmans},\ and\ \citenamefont {Clark}}]{Khatri2024a}%
  \BibitemOpen
  \bibfield  {author} {\bibinfo {author} {\bibfnamefont {N.}~\bibnamefont {Khatri}}, \bibinfo {author} {\bibfnamefont {G.}~\bibnamefont {Matos}}, \bibinfo {author} {\bibfnamefont {L.}~\bibnamefont {Coopmans}},\ and\ \bibinfo {author} {\bibfnamefont {S.}~\bibnamefont {Clark}},\ }\bibfield  {title} {\bibinfo {title} {{Quixer: A Quantum Transformer Model}},\ }\href {http://arxiv.org/abs/2406.04305} {\  (\bibinfo {year} {2024})},\ \Eprint {https://arxiv.org/abs/2406.04305} {arXiv:2406.04305} \BibitemShut {NoStop}%
\bibitem [{\citenamefont {Puig}\ \emph {et~al.}(2025)\citenamefont {Puig}, \citenamefont {Drudis}, \citenamefont {Thanasilp},\ and\ \citenamefont {Holmes}}]{Puig2025}%
  \BibitemOpen
  \bibfield  {author} {\bibinfo {author} {\bibfnamefont {R.}~\bibnamefont {Puig}}, \bibinfo {author} {\bibfnamefont {M.}~\bibnamefont {Drudis}}, \bibinfo {author} {\bibfnamefont {S.}~\bibnamefont {Thanasilp}},\ and\ \bibinfo {author} {\bibfnamefont {Z.}~\bibnamefont {Holmes}},\ }\bibfield  {title} {\bibinfo {title} {{Variational Quantum Simulation: A Case Study for Understanding Warm Starts}},\ }\href {https://doi.org/10.1103/PRXQuantum.6.010317} {\bibfield  {journal} {\bibinfo  {journal} {PRX Quantum}\ }\textbf {\bibinfo {volume} {6}},\ \bibinfo {pages} {010317} (\bibinfo {year} {2025})}\BibitemShut {NoStop}%
\bibitem [{\citenamefont {Cao}\ \emph {et~al.}(2025)\citenamefont {Cao}, \citenamefont {Jin},\ and\ \citenamefont {Liu}}]{Cao2025}%
  \BibitemOpen
  \bibfield  {author} {\bibinfo {author} {\bibfnamefont {Y.}~\bibnamefont {Cao}}, \bibinfo {author} {\bibfnamefont {S.}~\bibnamefont {Jin}},\ and\ \bibinfo {author} {\bibfnamefont {N.}~\bibnamefont {Liu}},\ }\bibfield  {title} {\bibinfo {title} {{Unifying framework for quantum simulation algorithms for time-dependent Hamiltonian dynamics}},\ }\href {https://doi.org/10.1103/fkh5-b669} {\bibfield  {journal} {\bibinfo  {journal} {Phys. Rev. Res.}\ }\textbf {\bibinfo {volume} {7}},\ \bibinfo {pages} {043186} (\bibinfo {year} {2025})}\BibitemShut {NoStop}%
\bibitem [{\citenamefont {Cerezo}\ \emph {et~al.}(2022)\citenamefont {Cerezo}, \citenamefont {Verdon}, \citenamefont {Huang}, \citenamefont {Cincio},\ and\ \citenamefont {Coles}}]{Cerezo2022}%
  \BibitemOpen
  \bibfield  {author} {\bibinfo {author} {\bibfnamefont {M.}~\bibnamefont {Cerezo}}, \bibinfo {author} {\bibfnamefont {G.}~\bibnamefont {Verdon}}, \bibinfo {author} {\bibfnamefont {H.-Y.}\ \bibnamefont {Huang}}, \bibinfo {author} {\bibfnamefont {L.}~\bibnamefont {Cincio}},\ and\ \bibinfo {author} {\bibfnamefont {P.~J.}\ \bibnamefont {Coles}},\ }\bibfield  {title} {\bibinfo {title} {{Challenges and opportunities in quantum machine learning}},\ }\href {https://doi.org/10.1038/s43588-022-00311-3} {\bibfield  {journal} {\bibinfo  {journal} {Nat. Comput. Sci.}\ }\textbf {\bibinfo {volume} {2}},\ \bibinfo {pages} {567} (\bibinfo {year} {2022})}\BibitemShut {NoStop}%
\bibitem [{\citenamefont {Mohseni}\ \emph {et~al.}(2022)\citenamefont {Mohseni}, \citenamefont {F{\"{o}}sel}, \citenamefont {Guo}, \citenamefont {Navarrete-Benlloch},\ and\ \citenamefont {Marquardt}}]{Mohseni2022}%
  \BibitemOpen
  \bibfield  {author} {\bibinfo {author} {\bibfnamefont {N.}~\bibnamefont {Mohseni}}, \bibinfo {author} {\bibfnamefont {T.}~\bibnamefont {F{\"{o}}sel}}, \bibinfo {author} {\bibfnamefont {L.}~\bibnamefont {Guo}}, \bibinfo {author} {\bibfnamefont {C.}~\bibnamefont {Navarrete-Benlloch}},\ and\ \bibinfo {author} {\bibfnamefont {F.}~\bibnamefont {Marquardt}},\ }\bibfield  {title} {\bibinfo {title} {{Deep Learning of Quantum Many-Body Dynamics via Random Driving}},\ }\href {https://doi.org/10.22331/q-2022-05-17-714} {\bibfield  {journal} {\bibinfo  {journal} {Quantum}\ }\textbf {\bibinfo {volume} {6}},\ \bibinfo {pages} {714} (\bibinfo {year} {2022})}\BibitemShut {NoStop}%
\bibitem [{\citenamefont {Rosati}\ and\ \citenamefont {Solana}(2024)}]{Rosati2023}%
  \BibitemOpen
  \bibfield  {author} {\bibinfo {author} {\bibfnamefont {M.}~\bibnamefont {Rosati}}\ and\ \bibinfo {author} {\bibfnamefont {A.}~\bibnamefont {Solana}},\ }\bibfield  {title} {\bibinfo {title} {{Joint-detection learning for optical communication at the quantum limit}},\ }\href {https://doi.org/10.1364/OPTICAQ.521637} {\bibfield  {journal} {\bibinfo  {journal} {Opt. Quantum}\ }\textbf {\bibinfo {volume} {2}},\ \bibinfo {pages} {390} (\bibinfo {year} {2024})},\ \Eprint {https://arxiv.org/abs/2312.13693} {arXiv:2312.13693} \BibitemShut {NoStop}%
\bibitem [{\citenamefont {Bilkis}\ \emph {et~al.}(2021)\citenamefont {Bilkis}, \citenamefont {Rosati},\ and\ \citenamefont {Calsamiglia}}]{Bilkis2021}%
  \BibitemOpen
  \bibfield  {author} {\bibinfo {author} {\bibfnamefont {M.}~\bibnamefont {Bilkis}}, \bibinfo {author} {\bibfnamefont {M.}~\bibnamefont {Rosati}},\ and\ \bibinfo {author} {\bibfnamefont {J.}~\bibnamefont {Calsamiglia}},\ }\bibfield  {title} {\bibinfo {title} {{Reinforcement-learning calibration of coherent-state receivers on variable-loss optical channels}},\ }in\ \href {https://doi.org/10.1109/ITW48936.2021.9611396} {\emph {\bibinfo {booktitle} {2021 IEEE Inf. Theory Work.}}}\ (\bibinfo  {publisher} {IEEE},\ \bibinfo {year} {2021})\ pp.\ \bibinfo {pages} {1--6}\BibitemShut {NoStop}%
\bibitem [{\citenamefont {Rosati}(2024)}]{Rosati2022}%
  \BibitemOpen
  \bibfield  {author} {\bibinfo {author} {\bibfnamefont {M.}~\bibnamefont {Rosati}},\ }\bibfield  {title} {\bibinfo {title} {{A learning theory for quantum photonic processors and beyond}},\ }\href {https://doi.org/10.22331/q-2024-08-08-1433} {\bibfield  {journal} {\bibinfo  {journal} {Quantum}\ }\textbf {\bibinfo {volume} {8}},\ \bibinfo {pages} {1433} (\bibinfo {year} {2024})},\ \Eprint {https://arxiv.org/abs/2209.03075} {arXiv:2209.03075} \BibitemShut {NoStop}%
\bibitem [{\citenamefont {Schuld}\ and\ \citenamefont {Petruccione}(2018{\natexlab{a}})}]{Schuld2018h}%
  \BibitemOpen
  \bibfield  {author} {\bibinfo {author} {\bibfnamefont {M.}~\bibnamefont {Schuld}}\ and\ \bibinfo {author} {\bibfnamefont {F.}~\bibnamefont {Petruccione}},\ }\href {https://doi.org/10.1007/978-3-319-96424-9} {\emph {\bibinfo {title} {{Supervised Learning with Quantum Computers}}}},\ Quantum Science and Technology\ (\bibinfo  {publisher} {Springer International Publishing},\ \bibinfo {address} {Cham},\ \bibinfo {year} {2018})\BibitemShut {NoStop}%
\bibitem [{Note1()}]{Note1}%
  \BibitemOpen
  \bibinfo {note} {These can be brought down to $\log k$ with binary encoding, but then requires additional comparator operations for the controlled gates; for an implementation with modest $k$, we favour the qubit over the gate overhead.}\BibitemShut {Stop}%
\bibitem [{\citenamefont {Thierrin}\ \emph {et~al.}(2022)\citenamefont {Thierrin}, \citenamefont {Alajaji},\ and\ \citenamefont {Linder}}]{Thierrin2022}%
  \BibitemOpen
  \bibfield  {author} {\bibinfo {author} {\bibfnamefont {F.~C.}\ \bibnamefont {Thierrin}}, \bibinfo {author} {\bibfnamefont {F.}~\bibnamefont {Alajaji}},\ and\ \bibinfo {author} {\bibfnamefont {T.}~\bibnamefont {Linder}},\ }\bibfield  {title} {\bibinfo {title} {{R{\'{e}}nyi Cross-Entropy Measures for Common Distributions and Processes with Memory}}\ }\href {https://doi.org/10.3390/e24101417} {10.3390/e24101417} (\bibinfo {year} {2022}),\ \Eprint {https://arxiv.org/abs/2208.06983} {arXiv:2208.06983} \BibitemShut {NoStop}%
\bibitem [{\citenamefont {LaRose}\ \emph {et~al.}(2019)\citenamefont {LaRose}, \citenamefont {Tikku}, \citenamefont {O'Neel-Judy}, \citenamefont {Cincio},\ and\ \citenamefont {Coles}}]{LaRose2019}%
  \BibitemOpen
  \bibfield  {author} {\bibinfo {author} {\bibfnamefont {R.}~\bibnamefont {LaRose}}, \bibinfo {author} {\bibfnamefont {A.}~\bibnamefont {Tikku}}, \bibinfo {author} {\bibfnamefont {{\'{E}}.}~\bibnamefont {O'Neel-Judy}}, \bibinfo {author} {\bibfnamefont {L.}~\bibnamefont {Cincio}},\ and\ \bibinfo {author} {\bibfnamefont {P.~J.}\ \bibnamefont {Coles}},\ }\bibfield  {title} {\bibinfo {title} {{Variational quantum state diagonalization}},\ }\href {https://doi.org/10.1038/s41534-019-0167-6} {\bibfield  {journal} {\bibinfo  {journal} {npj Quantum Inf.}\ }\textbf {\bibinfo {volume} {5}},\ \bibinfo {pages} {57} (\bibinfo {year} {2019})}\BibitemShut {NoStop}%
\bibitem [{\citenamefont {Schuld}\ and\ \citenamefont {Petruccione}(2018{\natexlab{b}})}]{Schuld2019}%
  \BibitemOpen
  \bibfield  {author} {\bibinfo {author} {\bibfnamefont {M.}~\bibnamefont {Schuld}}\ and\ \bibinfo {author} {\bibfnamefont {F.}~\bibnamefont {Petruccione}},\ }\href {https://doi.org/10.1007/978-3-319-96424-9} {\emph {\bibinfo {title} {{Supervised Learning with Quantum Computers}}}},\ Quantum Science and Technology\ (\bibinfo  {publisher} {Springer International Publishing},\ \bibinfo {address} {Cham},\ \bibinfo {year} {2018})\BibitemShut {NoStop}%
\bibitem [{\citenamefont {Mari}\ \emph {et~al.}(2021)\citenamefont {Mari}, \citenamefont {Bromley},\ and\ \citenamefont {Killoran}}]{Mari2021}%
  \BibitemOpen
  \bibfield  {author} {\bibinfo {author} {\bibfnamefont {A.}~\bibnamefont {Mari}}, \bibinfo {author} {\bibfnamefont {T.~R.}\ \bibnamefont {Bromley}},\ and\ \bibinfo {author} {\bibfnamefont {N.}~\bibnamefont {Killoran}},\ }\bibfield  {title} {\bibinfo {title} {{Estimating the gradient and higher-order derivatives on quantum hardware}},\ }\href {https://doi.org/10.1103/PhysRevA.103.012405} {\bibfield  {journal} {\bibinfo  {journal} {Phys. Rev. A}\ }\textbf {\bibinfo {volume} {103}},\ \bibinfo {pages} {012405} (\bibinfo {year} {2021})}\BibitemShut {NoStop}%
\end{thebibliography}%

\newpage

\onecolumngrid
\section{Setting and notation}
\label{sec:setting}

Let $D$ be the vocabulary size, $d$ the embedding dimension, $n=\log_2d$, $T$ the sequence
length, $c_T=\lceil\log_2T\rceil$, and $k$ the attention degree.  Let
$\{\ket{\bw^{(\ell)}}\}_{\ell=1}^{D}$ be the computational vocabulary basis and set
$A \eqdef E^{-1}:\C^d\to\C^D$.  We assume throughout that $ A$ and the classical
output map $ F:\C^d\to\C^D$ are isometries with the same range,
\begin{equation}
   A^\dagger A= F^\dagger F=\1_d,
  \qquad \operatorname{Ran} A=\operatorname{Ran} F.
\end{equation}
Hence
\begin{equation}
  G\eqdef A^\dagger F\in\C^{d\times d}
  \quad\text{is unitary}.%,\qquad   F= A G.
  \label{eq:Gfactor}
\end{equation}
As mentioned in the main text, this assumption can be dropped by embedding an arbitrary $G$ into a unitary on a larger space, via ancillary qubits, and introducing an explicit cirucit implementing $F$ at the branch-II readout. For simplicity, we do not discuss this case here, although most of the results we provide do generalize straightforwardly.

For the $\ell$-th word in the vocabulary define the corresponding embedded unnormalized token, its norm,
and its normalized version as
\begin{equation}
  \ket{\bx^{(\ell)}}\eqdef A^\dagger\ket{\bw^{(\ell)}},\qquad
  a_\ell\eqdef\|\bx^{(\ell)}\|^2,\qquad
  \ket{\bxi^{(\ell)}} \eqdef\frac{\ket{\bx^{(\ell)}}}{\sqrt{a_\ell}}.
  \label{eq:x-xi-def}
\end{equation}
We assume $a_\ell>0$ on the vocabulary's image through the embedding.  Since $ A$ is an
isometry,
\begin{equation}
  \sum_{\ell=1}^{D}\ket{\bx^{(\ell)}}\!\bra{\bx^{(\ell)}}=\1_d,
  \qquad 0\le a_\ell\le1,\qquad \sum_{\ell=1}^{D}a_\ell=\Tr[A^\dag A] = d.
  \label{eq:parseval-x}
\end{equation}
For a token sequence with vocabulary labels $\ell_i$ we abbreviate
$\bx_i\equiv \bx^{(\ell_i)}$, $\bxi_i\equiv\bxi^{(\ell_i)}$, and $a_i\equiv a_{\ell_i}$.

Our QSA uses only the normalized tokens $\ket{\bxi_i}$ to construct queries,
keys, and values.  The variational blocks $W,V$ are unitary and we define 
\begin{equation}
  s_{ij}\eqdef\bra{\bxi_j}W\ket{\bxi_i},\qquad
  g(s)\eqdef\sum_{p=0}^{k}c_p s^p,\qquad
  \lambda\eqdef\sum_{p=0}^{k} c_p,
\end{equation}
\begin{equation}
  \ket{\bz_j}\eqdef\sum_{i\le j}g(s_{ij})\,V\ket{\bxi_i},
  \qquad w_j\eqdef\|\bz_j\|_2.
  \label{eq:zdef}
\end{equation}
Instead, the \textit{prediction} uses the unnormalized tokens $\ket{\bx^{(\ell)}}$.  Including $G$ into the variational block $V$, the
normalized next-token distribution then is
\begin{equation}
  p_{j+1}^{(\ell)}
  \eqdef \frac{|\braket{\bx^{(\ell)}}{\bz_j}|^2}{\bw_j^2}
  =a_\ell\frac{|\braket{\bxi^{(\ell)}}{\bz_j}|^2}{\bw_j^2},
  \qquad \sum_{\ell=1}^{D}p_{j+1}^{(\ell)}=1,
  \label{eq:vocab-prob}
\end{equation}
where the normalization follows from \eqref{eq:parseval-x}.  This matches the classical expression derived in the main text after removing the feedforward non-linearity and assuming that $G$ is unitary, hence $F$ is an isometry. For the
observed target at step $j$ define
\begin{equation}
  p_j\eqdef p_{j+1}^{(\ell_{j+1})},\qquad
  f_j\eqdef\sqrt{p_j}=\frac{|\braket{\bx_{j+1}}{\bz_j}|}{w_j}\in[0,1],\qquad
  \phi_j\eqdef\arg\braket{\bx_{j+1}}{\bz_j}.
  \label{eq:wf}
\end{equation}
Thus every occurrence of $f_j^2$ below is literally the classical-model probability assigned
to the correct next token. 

For the attention, two kernel conventions are used:
\begin{itemize}[leftmargin=1.6em,itemsep=1pt]
\item The \textit{monomial} model corresponds to $c_p=\delta_{pk}$, i.e., $g(s)=s^k$ and $\lambda=1$;
the $P$-register being absent.
\item The \textit{polynomial} model corresponds to $c_p=\beta^p/p!$ with $\beta=\sqrt d$, i.e., $g(s)$ is
the degree-$k$ truncation of $e^{\beta s}$.  The absolute value of the inner product of weakly correlated unit vectors scales as
$|s|= O(d^{-1/2})$, so our choice $\beta=\sqrt d$ gives an $O(1)$ softmax-like argument.
\end{itemize}
We also set $\cN\eqdef|\{(i,j):i\le j\le T\}| = T(T+1)/2$.

Finally, the circuit uses the following registers
\begin{itemize}
\item $C_1$: $c_T$ qubits, controls prediction time-step $j$
\item $C_2$: $c_T$ qubits, controls token index $i$
\item $A$: $n$ qubits, holds values
\item $B_1,\dots,B_k$: $n$ qubits each, hold $k$ copies of query-key overlaps
\item $P$: $k$ qubits, controls monomial power
\item $R$: $1$ qubit, controls the target-POVM normalization.
\end{itemize}
The register $R$ is required only for target-dependent training readouts $\mu$, not the target-independent $\zeta$.  %No source or query normalization flags are needed, because the attention itself is defined on the normalized states $\bxi_i$.

\section{Derivation of the circuit output and observable expectations}
\label{sec:derivation}

Both $\mu, \zeta$ observables share the same preparation circuit, whose output $\ket{\psi_{\rm qsa}}$ implements attention in superposition; they differ only in the readout.

\subsection{Preparation}

\paragraph{Step 1 --- triangular control preparation.}
Apply a state-preparation unitary $\mathcal P$ such that
\begin{equation}
  \mathcal P\ket{0}_{C_1}\ket{0}_{C_2}
  =\frac{1}{\sqrt{\cN}}
   \sum_{j=1}^{T}\sum_{i\le j}\ket{j}\ket{i}
  \eqdef\ket{\triangle}.
  \label{eq:tri}
\end{equation}
One explicit construction is to first prepare the marginal distribution of
$j$,
\begin{equation}
  \ket{0}_{C_1}
  \longmapsto
  \sum_{j=1}^{T}\sqrt{\frac{j}{\cN}}\ket{j}_{C_1},
\end{equation}
and then, controlled on $j$, prepare a uniform superposition over the
$j$ admissible values of $i$,
\begin{equation}
  \ket{j}_{C_1}\ket{0}_{C_2}
  \longmapsto
  \ket{j}_{C_1}\frac{1}{\sqrt j}
  \sum_{i=1}^{j}\ket{i}_{C_2}.
\end{equation}
Both transformations are standard state-preparation operations.

\paragraph{Step 1$'$ --- polynomial preparation (polynomial model only).}
Apply a state-preparation unitary ${\rm PREP}$ such that
\begin{equation}
  {\rm PREP}\ket{0^k}_P
  =
  \sum_{p=0}^{k}
  \sqrt{\frac{c_p}{\lambda}}\,
  \ket{1^p0^{k-p}},
  \label{eq:prep}
\end{equation}
which is normalized because $\sum_{p=0}^{k}c_p/\lambda=1$. This is a standard preparation for a linear-combination-of-unitaries (LCU) algorithm that we employ to superpose QSA circuits with different monomial degrees. 
An explicit construction is given as follows, assuming $c_p\ge0$, defining the tail sums
\begin{equation}
  t_p\eqdef\sum_{q=p}^{k}c_q,
  \qquad
  t_0=\lambda,
\end{equation}
and the angles
\begin{equation}
  \theta_p
  =2\arcsin\sqrt{\frac{t_{p+1}}{t_p}},
  \qquad p=0,\ldots,k-1.
\end{equation}
Starting from $\ket{0^k}_P$, apply to qubit $P_{p+1}$ an
$R_y(\theta_p)$ rotation conditioned on the preceding $p$ qubits being
$\ket{1^p}$.  Thus the cascade continues from $p$ to $p+1$ unary
excitations with amplitude
$\sqrt{t_{p+1}/t_p}$, while it terminates at $p$ with amplitude
\begin{equation}
  \sqrt{1-\frac{t_{p+1}}{t_p}}
  =\sqrt{\frac{c_p}{t_p}}.
\end{equation}
Consequently the amplitude of $\ket{1^p0^{k-p}}$ telescopes to
\begin{align}
  \left(\prod_{q=0}^{p-1}
  \sqrt{\frac{t_{q+1}}{t_q}}\right)
  \sqrt{\frac{c_p}{t_p}}
  &=
  \sqrt{\frac{t_p}{t_0}}
  \sqrt{\frac{c_p}{t_p}}
  =
  \sqrt{\frac{c_p}{\lambda}},
\end{align}
with the $p=k$ branch obtained by continuing through all $k$ rotations.

\paragraph{Step 2 --- normalized controlled loads.}
Controlled on $C_2=\ket i$, load the normalized state $\ket{\bxi_i}$ into $A$ , and controlled additionally on $P=\ket{1^p 0^{k-p}}$, into each
$B_1^p$.  This can be done via amplitude-encoding unitaries $U_{\bxi_j}$ such that
$U_{\bxi_j}\ket{0^n}=\ket{\bxi_j}$. The degree-$p$ branch therefore contains
\begin{equation}
  \frac{1}{\sqrt{\cN}}\sum_{i\le j}\ket{j, i}_{C_{1,2}}\otimes
  \sum_{p=0}^{k}\sqrt{\tfrac{c_p}{\lambda}}\ket{1^p0^{k-p}}_P\otimes
  \ket{\bxi_i}_A\otimes(\ket{\bxi_i}^{\otimes p}\ket0^{\otimes(k-p)})_B.
\end{equation}

\paragraph{Step 3 --- variational blocks.}
Apply $V$ on $A$ and $W$ on each active $B_c$ register (controlled on $P$):
$\ket{\bxi_i}_A\mapsto V\ket{\bxi_i}$ and
$\ket{\bxi_i}_{B_c}\mapsto W\ket{\bxi_i}$.

\paragraph{Step 4 --- query uncomputation.}
Controlled on $C_1=\ket j$, apply $U_{\bxi_j}^\dagger$ to each active $B_c$ register (controlled on $P$).  Since
\begin{equation}
  \bra{0^n}U_{\bxi_j}^\dagger W\ket{\bxi_i}=\bra{\bxi_j}W\ket{\bxi_i}=s_{ij},
\end{equation}
the $\ket0_B$ component of branch $(j,i,p)$ acquires weight $s_{ij}^p$. The resulting state is
\begin{equation}
    \ket{\psi_{\rm qsa}} \eqdef \sum_{p,j,i}\sqrt{\frac{c_p}{\lambda \cN}}(\bra{\bxi_{j}}W\ket{\bxi_i})^p\, V\ket{\bxi_i}_A \otimes\ket{1^p 0^{k-p}, j,i}_{P C_{1,2}} 
\end{equation}
given in the main text.

\subsection{Branch-I readout}

\paragraph{Step 5 --- unitary readout and target POVM.}
The correct next token at each time-step is known during training, so we can implement the POVM effect $\dketbra{\bx_{j+1}}$.  Indeed, controlled on
$C_1=\ket j$, apply on the $R$ register the transformation
\begin{equation}
  \ket0_{R}\longmapsto
  \sqrt{a_{j+1}}\ket{0}_R+\sqrt{1-a_{j+1}}\ket{1}_R,
  \label{eq:Rt-filter}
\end{equation}
followed by $U_{\bxi_{j+1}}^\dagger$ on $A$.  The joint $R=0,A=0$ component acquires weight
\begin{equation}
  \sqrt{a_{j+1}}\,\bra{\bxi_{j+1}}V\ket{\bxi_i}
  =\bra{\bx_{j+1}}V\ket{\bxi_i}.
  \label{eq:target-povm-amp}
\end{equation}

\paragraph{Step 6 --- phase-alignment and uncomputation of step 1}
Apply the trainable phases $e^{\ii\varphi_j}$ on $C_1$, then $\cP^\dagger$ on $C_{1,2}$.

\paragraph{Step 6' --- uncomputation of step 1' (polynomial model only)}
Apply ${\rm PREP}^\dagger$.

\paragraph{Step 7 --- projection}
Finally, we perform an all-zero projection on all the remaining registers $A C_{1,2} P R$ with probability amplitude
\begin{equation}
    \cA = \frac{1}{\lambda \cN} \sum_{j=1}^T e^{\ii \varphi_j} \sum_{i=1}^j g(s_{ij}) \bra{\bx_{j+1}}V\ket{\bxi_i} = \frac{1}{\lambda \cN} \sum_{j=1}^T e^{\ii \varphi_j} \bra{\bx_{j+1}}V\ket{\bz_i} = \frac{1}{\lambda \cN} \sum_{j=1}^T  e^{\ii (\varphi_j + \phi_j)} w_j f_j.
\end{equation}
Therefore the expectation of this branch-I observable is
\begin{equation}
  \boxed{\;
  \mu =|\mathcal A|^2
  =\frac{1}{\lambda^2\cN^2}\left|\sum_{j=1}^T  e^{\ii (\varphi_j + \phi_j)} w_j f_j\right|^2,
  }
  \label{eq:mu-hybrid}
\end{equation}
as given in the main text.

\subsection{Branch-II readout}
Starting from $\ket{\psi_{\rm qsa}}$ after Step 4, we replace the following steps:

\paragraph{Step 5 --- uniform uncomputation of $C_2$.}
Instead of undoing the entire $\cP$ operation, we apply $H^{\otimes c_T}$ to uncompute $C_2$ with uniform weights on each branch. This yields a coefficient $\sqrt{T \cN}$ instead of $\cN$ in the final amplitude. 

\paragraph{Step 6' --- uncomputation of step 1' (polynomial model only).} Same as Step 5' in branch-I.

\paragraph{Step 7 --- discard and projection}
The resulting state of registers $C_1 A$ in the branch $\ket{0}_{C_2 P}$ is
\begin{equation}
    \frac{1}{\lambda \sqrt{T \cN}} \sum_{j=1}^T \ket{j}_{C_1} \otimes \ket{\bz_j}_A.
\end{equation}
In this case we perform an all-zero projection only on the registers $C_2 P$, discarding registers $A C_1$. Then the expectation of this branch-II observable is
\begin{equation}
  \boxed{\;\zeta =\frac{1}{\lambda^2T\cN}\sum_{j=1}^{T}\bw_j^2,\;}
  \label{eq:zeta}
\end{equation}
as given in the main text, which does not depend on the phases $\phi_j$.

\section{Properties of R\'enyi loss and relation to standard cross-entropy}
\label{sec:identities}
Let us first introduce a distribution induced by the prediction norms, that will be useful later.
\begin{definition}[Prediction-norm distribution]
\label{def:profiles}
On $j\in\{1,\dots,T\}$ we define the probability distribution
\begin{equation}
  q_j\eqdef\frac{w_j^{2}}{\sum_l w_l^{2}},
\end{equation}
quantifying the relative weight of each prediction $\ket{\bz_j}$ across different time-steps.
\end{definition}

We also define R\'enyi entropic quantities. In particular, see~\cite{Thierrin2022} for a discussion of possible definitions of R\'enyi cross-entropy; here we adopt the most straightforward generalization of standard cross-entropy.
\begin{definition}[R\'enyi divergence, entropy, cross-entropy]
\label{def:renyi}
For $\alpha\in(-\infty,1)\cup(1,\infty)$, a probability distribution $p$ and a (potentially subnormalized) distribution $r$, define the R\'enyi-$\alpha$ divergence (relative-entropy) and entropy as 
\begin{equation}
  D_\alpha(p\|r)\eqdef\frac{1}{\alpha-1}\log\sum_jp_j^{\alpha}r_j^{1-\alpha},
  \qquad
  H_\alpha(p)\eqdef\frac{1}{1-\alpha}\log\sum_jp_j^{\alpha}.
\end{equation}
and the R\'enyi-$\alpha$ cross-entropy as
\begin{equation}
  \boxed{\;{\rm CE}_\alpha(p\|r)\eqdef D_\alpha(p\|r)+H_\alpha(p)
  =\frac{1}{\alpha-1}\log\frac{\sum_jp_j^{\alpha}r_j^{1-\alpha}}{\sum_jp_j^{\alpha}}\;}
  \label{eq:cedef}
\end{equation}
In the $\alpha\to1$ limit $D_1=D_{\mathrm{KL}}$ is the Kullback-Leibler divergence, $H_1$ the Shannon entropy, and
${\rm CE}_1(p\|r)=-\sum_jp_j\log r_j$ the standard cross-entropy.  All defined quantities are non-negative.
\end{definition}
In particular, for $\alpha=1/2$,  we use repeatedly the following expressions:
\begin{equation}\label{eq:one_half_entropies}
  D_{1/2}(p||r) = -2\log\sum_j\sqrt{p_j r_j}, \qquad 
  H_{1/2}(p) = 2\log\sum_j\sqrt{p_j}, \qquad
  {\rm CE}_{1/2}(p\|r)=-2\log\frac{\sum_j\sqrt{p_j r_j}}{\sum_j\sqrt{p_j}}.
\end{equation}

\subsection{Entropic losses}
Here we show that the QSA training loss
\begin{equation}\label{eq:loss_qsa_main}
    \cL_{\rm qsa} := -\log\frac{(T+1) \mu}{2T \zeta}
\end{equation}
defined in the main text admits an insightful entropic formulation, under reasonable assumptions. For convenience, we restate the result presented in the main text using the notation introduced above:
\begin{theorem}(Thm.~1 of the main text)\label{thm:loss_entropic_sm}
    If $G$ is unitary and the trainable phases are aligned to the token-prediction overlaps, i.e., $\varphi_j=-\phi_j$ for all $j=1,\cdots,T$, then
    \begin{equation}\label{eq:loss_entropic_sm}
     \boxed{\;   \cL_{\rm qsa} = {\rm CE}_{1/2}(q||p) + D_{1/2}(u_T||q). \;}
    \end{equation}
\end{theorem}
Note that $p$ is not normalized across $j$, the time-step index, but rather across $\ell$, the vocabulary index. Here we interpret it as a subnormalized density across time-steps, which is sufficient for our purposes. An analogous use is made in standard generative tasks with the Shannon cross-entropy loss.

\begin{proof}
At phase-alignment and for $G$ unitary, inserting (\ref{eq:mu-hybrid},\ref{eq:zeta}) into \eqref{eq:loss_qsa_main} we have
\begin{equation}\label{eq:loss_substitute_mu_zeta}
    \cL_{\rm qsa} = -\log\left[\frac{T+1}{2T} \cdot \frac{T (\sum_j w_j f_j)^2}{\cN \sum_j w_j^2}\right] = -\log\frac{(\sum_j w_j f_j)^2}{T \sum_j w_j^2}.
\end{equation}
Inverting Definition~\ref{def:profiles}, write $w_j=\sqrt{q_j}\big(\sum_l w_l^{2}\big)^{1/2}$.  Then
$\sum_j w_jf_j=\big(\sum_l w_l^{2}\big)^{1/2}\sum_j\sqrt{q_j}f_j$, and substituting into \eqref{eq:loss_substitute_mu_zeta} we obtain
\begin{equation}\label{eq:loss_closed}
 \cL_{\rm qsa} = -\log \frac{(\sum_j \sqrt{q_j} f_j)^2}{T}.
\end{equation}
Now apply \eqref{eq:one_half_entropies} with $p=q$ and $r=f^{2}$, so that $\sqrt{p_jr_j}=\sqrt{q_j}f_j$:
\begin{equation}
  {\rm CE}_{1/2}(q\|f^{2})=-2\log\sum_j\sqrt{q_j}f_j+2\log\sum_j\sqrt{q_j},
  \label{eq:t1}
\end{equation}
and
\begin{equation}
  D_{1/2}(u_T\|q)=-2\log\sum_j\sqrt{(u_T)_jq_j}
  =-2\log\Big(\tfrac{1}{\sqrt T}\sum_j\sqrt{q_j}\Big)
  =\log T-2\log\sum_j\sqrt{q_j}.
  \label{eq:t2}
\end{equation}
Adding \eqref{eq:t1} and \eqref{eq:t2}, the terms $\pm2\log\sum_j\sqrt{q_j}$ cancel and the sum
equals \eqref{eq:loss_closed}. 
\end{proof}

We note in passing that the general expression \eqref{eq:loss_qsa_main} can also be interpreted as the fidelity between the superpositions of the anti-embedded predictions and that of the correct next words:
\begin{remark}\label{eq:loss_as_fidelity}[Fidelity interpretation of the loss]
At phase-alignment, define the normalized sequence states
\begin{equation}
 \ket{\Psi_{\rm pred}}
 =
 \sum_{j=1}^{T}e^{\ii \varphi_j}\sqrt{q_j}\,
 \ket{j}\otimes
 \frac{ F\ket{\bz_j}}{w_j},
 \qquad
 \ket{\Psi_{\rm corr}}
 =
 \frac{1}{\sqrt T}
 \sum_{j=1}^{T}
 \ket{j}\ket{\bw^{(\ell_{j+1})}} .
\end{equation}
Their fidelity is
\begin{equation}
 F
 =
 |\langle\Psi_{\rm pred}|\Psi_{\rm corr}\rangle|^2
 =
 \frac1T
 \left(\sum_{j=1}^{T}\sqrt{q_jp_j}\right)^2
 =
 \frac{T+1}{2T}\frac{\mu}{\zeta},
\end{equation}
so that $\mathcal L_{\rm qsa}=-\log F$.  
\end{remark}
Thus the order-$1/2$ structure of the
loss has a simple amplitude-level origin: the Bhattacharyya-type sum
$\sum_j\sqrt{q_jp_j}$ is precisely the overlap amplitude of the prediction- and correct-state superpositions.  This interpretation is not required for the entropic
decomposition above, but provides a useful geometric meaning.

From \eqref{eq:loss_entropic_sm}, defining the standard cross-entropy loss
\begin{equation}
    \cL_1 \eqdef {\rm CE}_1(u_T||p),
\end{equation}
we can study the relation between the two losses, as mentioned in the main text. In particular, this is based on the key observation that $\cL_{\rm qsa}$ can be written directly as a R\'enyi-$1/2$ relative-entropy between the uniform distribution and a subnormalized distribution proportional to the correct-next-token prediction probabilities $p_j$. 
\begin{corollary} (Corollary~1 of the main text)\label{cor:loss_zero}
        Under the assumptions of Thm.~\ref{thm:loss_entropic_sm} , it holds $\cL_{\rm qsa} \leq \cL_1 + D_1(u_T||q)$, which is saturated at $\cL_{\rm qsa}=0 \Rightarrow \cL_1=0$, the converse being false.
\end{corollary}
\begin{proof}
Define the non-negative, generally subnormalized success measure
\begin{equation}
  r_j\eqdef q_jf_j^2 .
\end{equation}
Then, directly from the definition of the R\'enyi divergence,
\begin{align}
  D_{1/2}(u_T\|r) =-2\log\sum_j\sqrt{\frac1T q_jf_j^2} 
  =\log T-2\log\sum_j\sqrt{q_j}f_j
   \equiv\mathcal L_{\rm qsa} .
  \label{eq:LB-single-renyi}
\end{align}
Although $r$ need not be normalized, R\'enyi-order monotonicity still applies:
writing $Z=\sum_jr_j$ and $\widehat r=r/Z$ for the normalized distribution, we have
\begin{equation}
  D_\alpha(u_T\|r)
  =D_\alpha(u_T\|\widehat r)-\log Z,
\end{equation}
and the second term is independent of $\alpha$.  Hence by monotonicity of the R\'enyi divergence in $\alpha$ we obtain
\begin{equation}
  \mathcal L_{\rm qsa}
  =D_{1/2}(u_T\|\hat r) -\log Z
  \le D_1(u_T\|\hat r) -\log Z.
\end{equation}
Finally,
\begin{align}
  D_1(u_T\|r)
  =\frac1T\sum_j\log\frac{1/T}{q_jf_j^2} 
  ={\rm CE}_1(u_T\|p)+ D_1(u_T\|q),
\end{align}
which proves the Corollary's inequality. 

As for the zeros, we have that both terms in $\cL_{\rm qsa}$ are non-negative, and $0\le p_j\le1$ are subnormalized, therefore
\begin{equation}
  \boxed{\;
  \mathcal L_{\rm qsa}=0
  \iff
  p_j=1\ \text{for all }j
  \quad\text{and}\quad
  q=u_T .
  \;}
\end{equation}
For similar reasons, $\cL_1$ vanishes iff $p_j=1$ for all $j$.  Hence
\begin{equation}
  \boxed{\;
  \mathcal L_B=0
  \Longrightarrow
  \mathcal L_1=0 .
  \;}
\end{equation}
Thus the coherent loss imposes the same perfect-prediction condition as the
uniform Shannon loss, together with the additional requirement of a uniform
norm profile.  Furthermore, at this common ideal point the bound
$\mathcal L_B\le
\mathcal L_1^{\rm unif}+D_1(u_T\|q)$ is saturated.
\end{proof}
As a consequence, $\cL_1$ can be used at testing stage to compare our QSA with models employing the standard cross-entropy, avoiding distinct-target issues.

\subsection{Excess losses}\label{sec:excess-losses}

As mentioned in the main text, we can show that the losses are lower-bounded by quantities dependent on the embedded tokens' squared norms; defining the normalized-correct-token prediction probabilities $\rho_j := p_j/a_{j+1}\geq 1$, and taking the relative-entropy-term minimizer $q=u_T$ we have
\begin{equation}
    \cL_{\rm qsa} \geq{\rm CE}_{1/2}(u_T||p) = -2\log \sum_j \frac1T \sqrt{\tilde p_j a_{j+1}} \geq - \log a_A,
\end{equation}
and similarly
\begin{equation}
    \cL_1 \geq-\frac1T\sum_j\log a_{j+1}=-\log a_G,
\end{equation}
 where
\begin{equation}
  a_{A}\eqdef\frac1T\sum_{j=1}^Ta_{j+1},\qquad
  a_{G}\eqdef
\prod_{j=1}^T (a_{j+1})^{1/T},
  \label{eq:alpha-means}
\end{equation}
are the arithmetic and geometric means of the tokens' squared norms. Since $\sum_j a_{j+1} = d$ from the word normalization constraint in vocabulary space, we expect that $a_{R,G}<1$ and therefore the zero-point of Corollary~\ref{cor:loss_zero} is in general not achievable.

This is not in itself an issue, as much as a feature that identifies $\cL_{\rm qsa, 1}$ as good functions for optimizing also the embedding  via their norm-dependence. Indeed, using $\cL_{\rm qsa}$ lets training reallocate leverage from
less useful vocabulary directions toward words that matter under the data distribution.

Still, when testing the trained model's performance we would like to employ loss functions with a clear zero floor, particularly to check that the joint-minima conditions of Corollary~\ref{cor:loss_zero} are met. 
To this purpose, we define the excess losses
\begin{equation}\label{eq:loss_excess}
    \cL_{\rm qsa}^{\rm ex} \eqdef \cL_{\rm qsa} + \log a_A, \qquad \cL_1^{\rm ex} \eqdef \cL_1 + \log a_G.
\end{equation}
for which similar bounds and joint-minimization conditions are valid, this time truly achievable by the model.

\begin{corollary}(Corollary~2 of the main text)
\label{prop:excess-decomposition}
    Under the assumptions of Thm.~\ref{thm:loss_entropic_sm} , it holds $\cL_{\rm qsa}^{\rm ex} \leq \cL_1^{\rm ex} + D_1(u_T||q) + D_1(u_T||a)$, and $\cL_{\rm qsa}^{\rm ex}=0 \Rightarrow \cL_1^{\rm ex}=0$, the converse being false. For a balanced frame, $a=u_T$, the bound reduces to
$\cL_{\rm qsa}^{\rm ex}\le\cL_1^{\rm ex}+D_1(u_T\|q)$ and is saturated at
the common zero $q=a=u_T$.
\end{corollary}

\begin{proof}
Introducing the normalized capacity profile
\begin{equation}
  \tilde a_j=\frac{a_{j+1}}{Ta_A},
\end{equation}
We have
\begin{equation}
  D_1(u_T\|\tilde a)
  =-\log T-\frac1T\sum_j\log \tilde a_j
  =\log\frac{a_A}{a_G}.
  \label{eq:capacity-gap}
\end{equation}
Adding $\log a_A$ to the bound of Corollary~\ref{cor:loss_zero} immediately yields the desired bound.

For the zero set, using $\cL_{\rm qsa}^{\rm ex} = -\log(F/a_A)$ with
$F=T^{-1}(\sum_j\sqrt{q_jp_j})^2$ gives
\begin{equation}
  \frac{F}{a_A}
  =
  \left(\sum_j\sqrt{\tilde a_{j} q_j\rho_j}\right)^2
  \le
  \left(\sum_j\sqrt{\tilde a_{j} q_j}\right)^2
  \le1 ,
  \label{eq:excess-zero-proof}
\end{equation}
where the second inequality is Cauchy--Schwarz.  Equality holds iff
\begin{equation}
  \rho_j=1\quad\forall j,
  \qquad q=\tilde a.
\end{equation}
Hence
\begin{equation}
 \boxed{\; \cL_{\rm qsa}^{\rm ex}=0 
  \iff \rho_j=1\ \forall j\ \text{and}\ q=\tilde a.\;}
\end{equation}

On the other hand, we can write
\begin{equation}
    \cL_1^{\rm ex} = -\frac1T\sum_j \log(\rho_j a_{j+1}) + \log a_G
  =-\frac1T\sum_j\log\rho_j ,
\end{equation}
and therefore
\begin{equation}
\boxed{\;  \cL_1^{\rm ex}=0
  \iff \rho_j=1\quad\forall j.\;}
\end{equation}
This proves the implication; the converse fails because
$\cL_1^{\rm ex}=0$ places no constraint on $q$. Clearly, at the zero condition the additional term $D_1(u_T||\tilde a)$ in the inequality does not vanish unless the token-norms distribution is uniform. 
\end{proof}

\section{Optimality of phase-alignment}
\label{part:phase}
Here we prove that the phase-alignment point $\varphi_j=-\phi_j$ is a critical point of $\cL_{\rm qsa}$; furthermore, at fixed values of all the other parameters, the point is also a minimum of the loss. 

Let $\bm\theta$ collect all non-phase trainable parameters (in particular those of $W,V,E, F$), and
define
\begin{equation}
  A_j(\bm\theta)\eqdef\ip{\bx_{j+1}}{\bz_j}
  =r_j(\bm\theta)e^{\ii\phi_j(\bm\theta)},\qquad r_j(\bm\theta)>0,
  \qquad j=1,\dots,T.
\end{equation}
Terms with $A_j=0$ carry no phase information and may simply be omitted from the active phase
set, so we assume $r_j>0$ below.  The branch-I, II readouts and the training loss can be written as
\begin{equation}
  \mu(\bm\varphi,\bm\theta)=\frac{1}{\lambda^{2}\cN^{2}}
  \Big|\sum_{j=1}^{T}e^{\ii\varphi_j}A_j(\bm\theta)\Big|^{2},\qquad \zeta(\bm\theta)=\frac{1}{\lambda^2 T\cN}\sum_j w_j(\bm\theta)^2,\qquad
    \mathcal L_{\rm qsa}(\bm\varphi,\bm\theta) =-\log\!\left[\frac{T+1}{2T}\frac{\mu(\bm\varphi,\bm\theta)}{\zeta(\bm\theta)}\right].
  \label{eq:muphi}
\end{equation}
Since the branch-II observable is independent of $\bm\varphi$, the phase-optimization landscape of
$\mathcal L_{\rm qsa}$ is exactly that of $-\log\mu$.  Moreover, $\mu$ is invariant under the global
shift $\bm\varphi\mapsto\bm\varphi+\alpha\mathbf 1$, with $\alpha$ a global phase-gauge parameter and $\mathbf 1$ the all-one vector.% that we can consider fixed and does not influence our results below. 

\begin{theorem}[Phase alignment, convergence, and profiling]
\label{thm:phase}
Fix $\bm\theta$ with $r_j(\bm\theta)>0$ for all active terms, and define the set of gauge-equivalent solutions
\begin{equation}
  \Phi^\star(\bm\theta)
  \eqdef\left\{\bm\varphi:\ \varphi_j=-\phi_j(\bm\theta)+\alpha,
  \ \alpha\in\R\right\}.
\end{equation}
Then the following statements hold:

\begin{enumerate}[label=(\alph*),leftmargin=2em,itemsep=3pt]
\item Exact global phase optimum.
For every $\bm\varphi$,
\begin{equation}
  \mu(\bm\varphi,\bm\theta)\le
  \mu_{\max}(\bm\theta)
  \eqdef\frac{1}{\lambda^{2}\Ntri^{2}}
  \left(\sum_j r_j(\bm\theta)\right)^2,
  \label{eq:mu-max-phase}
\end{equation}
with equality if and only if $\bm\varphi\in\Phi^\star(\bm\theta)$.  Thus, modulo the global
gauge, phase alignment is the unique global minimizer of $\mathcal L_{\rm qsa}$.

\item Landscape and local stability.
Every finite-loss critical point has each term $e^{\ii\varphi_j}A_j$ either aligned or
anti-aligned with their sum.  The aligned point is the unique minimum (up to gauge);
every other finite-loss critical point is a strict saddle of $\mathcal L_B$.  At alignment,
with
\begin{equation}
  P\eqdef\sum_j r_j ,\qquad \pi_j\eqdef r_j/P,
\end{equation}
the phase Hessian is
\begin{equation}
  K_{jl}\eqdef
  \left[\nabla^2_{\bm\varphi}\mathcal L_{\rm qsa}\right]_{jl}
  \Big|_{\Phi^\star}
  =2(\pi_j\delta_{jl}-\pi_j\pi_l),
  \label{eq:phase-hessian}
\end{equation}
so that
\begin{equation}
  v^\top Kv
  =2\sum_j\pi_jv_j^2-2\left(\sum_j\pi_jv_j\right)^2
  =2\Var_{\pi}(v)\ge0.
  \label{eq:phase-hessian-var}
\end{equation}
Its only zero mode is the global-gauge direction $v=\mathbf 1$; hence $K$ is strictly positive-definite on the quotient space.
\end{enumerate}
\end{theorem}

\begin{proof}
\textit{(a)}  The triangle inequality gives
\begin{equation}
 \left|\sum_j e^{\ii\varphi_j}A_j\right|\le\sum_j|A_j|.
\end{equation}
Equality holds exactly when all summands have a common phase, i.e.,
$\varphi_j+\phi_j=\alpha$ for every $j$.  This proves both the bound and uniqueness modulo the
global gauge.

\textit{(b)}  Let $S\eqdef\sum_l e^{\ii\varphi_l}A_l$.  At finite loss $S\ne0$, and, using $\partial_{\varphi_j}S = i e^{\ii \varphi_j} A_j$, we have
\begin{equation}\label{eq:first_der_mu}
  \partial_{\varphi_j}\mu
 = -\frac{2}{(\lambda \cN)^2} {\rm Im}\!\left(e^{\ii\varphi_j}A_j  S^*\right) \propto|A_j|\,|S|\sin(\varphi_j+\phi_j-\arg S)
\end{equation}
Thus every critical term satisfies $\varphi_j+\phi_j-\arg S\in\{0,\pi\}$.  The all-$0$ fully aligned 
configuration is precisely $\Phi^\star$.  If instead some term $j$ is anti-aligned, write
$R_j=\sum_{l\ne j}e^{\ii\varphi_l}A_l$.  At that critical point $R_j$ is aligned to $S$ and
$e^{\ii\varphi_j}A_j$ is anti-aligned with $R_j$; writing
\begin{equation}
    \mu \propto \big|e^{\ii\varphi_j}A_j+R_j\big|^{2}
=|A_j|^{2}+|R_j|^{2}+2|A_j|\, |R_j| \cos(\varphi_j+\phi_j-\arg S),
\end{equation}
it is clear that varying only $\varphi_j$ moves away
from a minimum of the cosine term
and strictly increases $\mu$.  Hence
$\mathcal L_{\rm qsa}$ has a negative-curvature direction and the point is a strict saddle.

At alignment, we have $\arg S=0$ and
\begin{equation}
    \mu|_{\bm \varphi^*} = \frac{P^2}{(\lambda\cN)^2}, \qquad \partial_{\varphi_j}\mu|_{\bm\varphi^*} = 0. 
\end{equation}
Furthermore, deriving again \eqref{eq:first_der_mu}
\begin{equation}
   \partial_{\varphi_\ell} \partial_{\varphi_j}\mu = \frac{2}{(\lambda\cN)^2} \left[{\rm Re}(e^{\ii(\varphi_j-\varphi_\ell)}A_j A_\ell^*) - \delta_{\ell,j}\, {\rm Re}(e^{\ii \varphi_j}A_j S^*) \right]
\end{equation}
and therefore
\begin{equation}
    \partial_{\varphi_\ell} \partial_{\varphi_j}\mu|_{\bm\varphi^*} = \frac{2 P^2}{(\lambda\cN)^2} (\pi_\ell \pi_j -\delta_{\ell,j}\, \pi_j). 
\end{equation}
Then, differentiating
\begin{equation}
    \partial_{\varphi_\ell}\partial_{\varphi_j}(-\log\mu) = \frac{1}{\mu^2}\partial_{\varphi_\ell}\mu \partial_{\varphi_j}\mu - \frac{1}{\mu} \partial_{\varphi_\ell}\partial_{\varphi_j}\mu
\end{equation}
and substituting the values at phase-alignment gives Eq.~\eqref{eq:phase-hessian}.  Equation
\eqref{eq:phase-hessian-var} is the weighted variance of $v$ under $\pi$; because every
$\pi_j>0$, it vanishes exactly when $v$ is constant, which is precisely the
global phase gauge, so $K$ is positive-definite on the quotient.
\end{proof}
Theorem~\ref{thm:phase} shows that, for every fixed $\bm\theta$, the aligned
orbit $\Phi^\star(\bm\theta)$ is the unique minimum of the phase landscape,
while every other finite-loss critical point is a strict saddle.  Thus the
phase variables introduce no spurious local minima.  As a consequence, gradient-based optimization methods are expected to converge to the aligned orbit from generic initialization points with $\mu>0$.

Moreover, since the phase Hessian is positive definite on the gauge-direction-quotient at
$\Phi^\star(\bm\theta)$ for every fixed $\bm\theta$, around the phase-alignment one always finds a local minimum, i.e.,
\begin{equation}
  \mathcal L_{\rm qsa}
  (\bm\varphi^\star+\delta\bm\varphi,\bm\theta)
  =
  \mathcal L_{\rm qsa}
  (\bm\varphi^\star,\bm\theta)
  +\frac12\,
  \delta\bm\varphi^\top K\,\delta\bm\varphi
  \geq \mathcal L_{\rm qsa}
  (\bm\varphi^\star,\bm\theta).
\end{equation}
up to $o(\|\delta\bm\varphi\|^2)$. Hence, locally, gradient descent in the phase coordinates drives deviations
back toward the instantaneous aligned values
$\varphi_j^\star=-\phi_j$ modulo the global phase gauge. This ensures that an alternating optimization of $\bm\varphi$ and $\bm\theta$ will converge to phase-alignment for the optimal $\bm \theta$.
In the numerics, for simplicity we optimize all variables jointly, showing empirical convergence that is not covered by the present analytical results.

\section{End-to-end training and sampling complexity}
\label{part:cx}
Here we study in detail the end-to-end training complexity of our QSA circuit. We then compare it with a benchmark relying on an efficient classical implementation of the polynomial kernel, providing conditions for an training-advantage regime. Finally, we perform a similar analysis for the sampling complexity, i.e., the cost of generating new predictions once trained, by sampling from the trained model distribution. 

\subsection{Training complexity}
\subsubsection{Quantum self-attention}
In terms of hardware size, our QSA circuit requires a total number of qubit registers 
\begin{equation}
  n_{\mathrm{tot}}=
  \underbrace{(k+1)\log d}_{A,B_1^k}
  +\underbrace{2\lceil\log T\rceil}_{C_1,C_2}
  +\underbrace{k}_{P}
  +\underbrace{1}_{R}.
  \label{eq:qubits-hybrid}
\end{equation}
We note that register $R$ is required only for branch-I readout, while the size of register $P$ can be brought down to $\log k$ employing a binary encoding of the polynomial terms' degree, which however increases the gate complexity of the preparation step $\cP$. Since $k$ is fixed and can be kept relatively small for the approximation of softmax to work, here we favour a qubit overhead over a gate one. 

The gate count, determining the bare cost of implementing an instance of the circuit, is provided below. The cost is equal for both polynomial and monomial kernels at the same degree. We include explicitly the cost of amplitude-encoding of the normalized tokens $\ket{\bxi_i}=U_i\ket{0}$, employing a conservative estimate \textit{in the absence of QRAM}. This provides the leading cost in $d$, $k$, and $T$. Note that we do not incldue the cost of processing and training the embedding, as this is performed on a classical computer and it is shared by all quantum and classical models that we compare. 
\begin{lemma}[QSA Gate count]
\label{prop:gates}
For either kernel, assuming an $L$-layer ansatz cost of $C_{\rm ansatz} = O(L\log d)$ where $L$ on $\log d$ qubits, the total gate count of QSA preparation and readout circuits is, at leading order,
\begin{align}
 C_{\mathrm{circ}}=\Ord{kTd}.
 \label{eq:gates-hybrid}
\end{align}
\end{lemma}

\begin{proof}
The exact preparation of an arbitrary normalized $d$-dimensional state $\ket{\bxi}$ via QRAM-unassisted amplitude-encoding costs $\Ord d$ gates.  This is applied in parallel to $k+1$ registers and then, controlled by $C_1$, in superposition to $T$ distinct timesteps, resulting in a total complexity $\Ord{kTd}$. A similar cost applies to the controlled-uncomputation on registers $B_1^k$. Additionally, the uncomputation on register $A$ for branch-I readout, with an additional control value on register $R$, costs a subleading $\Ord{Td}$, and the $C_1$-controlled preparation of the norm-dependent state on register $R$ for implementing the POVM element $\dketbra{\bx}$ costs $O(T)$.

The cost of explicitly implementing $G$ and $F$ would be analogous but we assume, as in the rest of the paper, that they are unitary/isometric and thus implicitly implemented via $V, W$. Defining $C_{\rm ansatz}(n)$ the gate-count for a single ansatz on an $n$-qubit register, the total cost of $V\otimes W^{\otimes k}$ is $O(k C_{\rm ansatz}(n))$.

On the controls, the (un)-preparations on registers $C_{1,2}$ cost $O({\rm poly}\log T)$ since they require to construct superpositions of at most $T(T+1)/2$ terms on $2\log T$ qubits. Instead, given the unary encoding employed on register $P$, its (un)-preparation in a superposition of $k$ terms on $k$ qubits costs $O(k)$. 

Finally, the extra variational correcting phases on $C_1$ cost $O(T)$. 

The resulting leading gate count is thus as stated in the main text. 
\end{proof}

In order to calculate the end-to-end training complexity, we now evaluate the additional number of shots, i.e., calls to the circuit with bare complexity \eqref{eq:gates-hybrid}, needed to estimate precisely the two readout observables:
\begin{lemma}[Shot cost]
\label{cor:budget}
For the estimation $\mathcal L_{\rm qsa}$ with additive error $\epsilon$, a sufficient total number of samples of the branch-I,II circuits is
\begin{equation}
N_{\rm shots}=O(\epsilon^{-2} \mu^{-1}).
 \label{eq:NB-hybrid}
\end{equation}
Similarly $N_{\rm shots}^{\rm grad}= O(k C_{\rm ansatz} N_{\rm shots})$ for estimating its gradient via the parameter-shift rule~\cite{Schuld2019,Mari2021}, with the standard assumption that the single-shot observable variance depends only weakly on the parameter shift.
\end{lemma}
\begin{proof}
    Let $p\in(0,1]$ be a projector expectation estimated from $N$ i.i.d. Bernoulli repetitions.  Then the
relative error obeys
\begin{equation}
 \epsilon=\sqrt{\frac{1-p}{Np}}\le\frac1{\sqrt{Np}}\Rightarrow
 \qquad N\ge\frac1{p\epsilon^2}.
\end{equation}
Using the fidelity interpretation of the loss, Remark~\ref{eq:loss_as_fidelity}, we have
\begin{equation}
  \mu=\frac{2T}{T+1}\zeta F<2\zeta.
  \label{eq:mu_zeta_bound}
\end{equation}
Hence the contribution associated with $\zeta$ can exceed that associated
with $\mu$ by at most a constant factor, and the asymptotic shot complexity
is controlled by $\mu$.  Indeed, allocating any fixed nonzero fractions of
$N_{\rm shots}$ to the branch-I and branch-II observables gives
\begin{equation}
  \epsilon^2
  \leq
  \frac{1}{f N_{\rm shots}}
  \left(
    \frac{1}{\mu}+\frac{1}{\zeta}
  \right)
  =
  O\!\left(\frac{1}{N_{\rm shots}\mu}\right),
\end{equation}
where in the last equality we used Eq.~\eqref{eq:mu_zeta_bound}, and inverting we obtain the desired statement for estimating the ratio $\mu/\zeta$ with relative error $\epsilon$. The same scaling is valid for estimating $\cL_{\rm qsa}=-\log(\mu/\zeta) + {\rm const.}$ to \textit{additive} error $\epsilon$. 

Finally, the gradients can be estimated via the parameter-shift rule, which, under standard assumptions, multiplies the complexity by $O(k C_{\rm ansatz})$. We note that in the non-standard case where the shifted observables are not of the same order of the original, i.e., $\mu_{\pm}\neq O(\mu)$, then the observable-dependence in $N_{\rm shots}^{\rm grad}$ shifts to $\mu^{-2}$.
\end{proof}
We thus obtain the result of the main text:

\begin{theorem}
\label{prop:qcost}
The end-to-end QSA training complexity with additive error $\epsilon$ on $\cL_{\rm qsa}$ is, to leading order
\begin{equation}
\boxed{ C_{k-\rm qsa} = O\left(\frac{k^2Td C_{\rm ansatz}}{\epsilon^2\mu}\right)\;}
 \label{eq:qcost-hybrid}
\end{equation}
\end{theorem}
\begin{proof}
 The end-to-end-training complexity for an optimizer step is given by $C_{\rm circ}\cdot N_{\rm shots}^{\rm grad}$, taking into account shots for estimating both the loss and its gradient. Combining Lemmas~\ref{prop:gates}, \ref{cor:budget} we obtain the desired statement.
\end{proof}

\subsubsection{Polynomial-kernel classical self-attention}
As mentioned in the main text, implementing standard softmax-kernel classical self-attention (nl-CSA) has a dominant scaling $O(d T^2)$~\cite{Vaswani}, which is quadratically worse in $T$ with respect to our end-to-end QSA complexity estimate  \eqref{eq:qcost-hybrid}. However, our QSA employs a polynomial kernel, for which a classical equivalent can be implemented at a different cost than softmax. We now evaluate the cost of such a polynomial-kernel classical self-attention (CSA) baseline. For simplicity, we stick to the ket-bra notation even though we are considering classical vectors and use normalized tokens;; the same results hold for normalized ones.

We start by introducing a convenient map for the polynomial kernel:
\begin{proposition}[Degree-$k$ feature map]
\label{prop:feature}
Let $m_k\eqdef\binom{d+k-1}{k}$ be the dimension of the space of homogeneous degree-$k$
polynomials in $d$ variables. Then the map $\varphi_k:\C^d\to\C^{m_k}$,
$\varphi_k(\bxi)_{\bm\alpha}=\sqrt{\binom{k}{\bm\alpha}}\,\prod_{i=1}^d x_i^{\alpha_i}$ over multi-indices
$|\bm\alpha|=k$ satisfies
\begin{equation}
  \ip{\varphi_k(\bxi_i)}{\varphi_k(\bxi_j)}=\ip{\bxi_i}{\bxi_j}^{k}
  \qquad\text{for all } \bxi_i,\bxi_j .
\end{equation}
For the inhomogeneous kernel $g(s)=\sum_{p\le k}c_ps^{p}$ one concatenates the maps
$\sqrt{c_p}\,\varphi_p$ for $p=0,\dots,k$, of total dimension
$\sum_{p\le k}m_p=\binom{d+k}{k}=O(m_k)$.
\end{proposition}
\begin{proof}
    The result follows simply by expanding $\braket{\bxi_i}{\bxi_j}^k$ via the multinomial theorem. 
\end{proof}

Using this kernel, one can speed-up the calculation of polynomial self-attention:
\begin{theorem}[Classical cost $k$-CSA]
\label{prop:ccost}
The polynomial-kernel self-attention prediction
\begin{equation}
 \ket{\bz_j}=\sum_{i\le j}g\big(\ip{\bxi_j}{W\bxi_i}\big)V\ket{\bxi_i}
\end{equation}
is computable for all $j$ in
\begin{equation}\label{eq:classical_cost_final}
\boxed{C_{k-\rm csa}=\Ord{T d m_k}\;}
\end{equation}
arithmetic operations, without forming the $T\times T$ attention matrix.
\end{theorem}
\begin{proof}
Using Proposition~\ref{prop:feature},
$g(\ip{\bxi_j}{W\bxi_i})=\ip{\varphi(\bxi_j)}{\varphi(W\bxi_i)}$. Hence
\begin{equation}\label{eq:pred_k_csa}
  \ket{ \bz_j}=\sum_{i\le j}\ket{V\bxi_i}\braket{\varphi(\bxi_i)}{ \varphi(W\bxi_j)}
  = S_j\ket{\varphi(W\bxi_j)},
\end{equation}
where we have defined $S_j = \sum_{i\le j}\ketbra{V\bxi_i}{\varphi(\bxi_i)}$. The calculation of ${S_j}_{j=1}^T$ can be done recursively with rank-one updates $S_j=S_{j-1}+\ketbra{V\bxi_j}{\varphi(\bxi_j)}$, costing
$\Ord{m_kd}$ operations for each $j=1,\cdots,T$, which gives the scaling \eqref{eq:classical_cost_final}.  The matrix-vector products \eqref{eq:pred_k_csa} require a further subleading $O(m_k)$ operations per time-step.
\end{proof}
As a consequence, $k$-CSA is cheaper than nl-CSA in the long-sequence regime $T\ge m_k$, while at the same time providing a good approximation of the softmax kernel for sufficiently large degree $k$.

\subsubsection{Advantage threshold}
\begin{theorem}[Advantage threshold]
\label{thm:advantage}
Let $\mu$ denote the joint branch-I probability of the hybrid circuit, including the target
POVM.  At fixed precision and number of ansatz layers $\epsilon, L=O(1)$, the QSA end-to-end training complexity is cheaper
than the cheapest CSA evaluation of the same normalized-token attention function iff
\begin{equation}
 \boxed{\;\mu\geq O\left(\frac{k^2 \log d}{\min\{T,m_k\}}\right),\qquad
 m_k=\binom{d+k-1}{k}.\;}
 \label{eq:advantage-hybrid}
\end{equation}
\end{theorem}

\begin{proof}
Combine \eqref{eq:qcost-hybrid} with
$C_{\mathrm{cl}}=\Ord{Td\min\{m_k,T\}}$ and solve for $\mu$, hiding $O(1)$ contributions.
\end{proof}

\section{Extrapolation to the advantage regime}
The empirical behaviour of $\mu$ after training is well-described by a saturating curve for large $T$. To better study the $d,T$-dependence we write
\begin{equation}
\mu(T,d)=\muinf(d)\bigl[1+\varepsilon(T,d)\bigr].
\label{eq:excessdef}
\end{equation}
All fits are realized in $\log$-space with weighted nonlinear least squares,
\begin{equation}
\hat\vartheta=\arg\min_\vartheta\sum_{i}
\frac{\bigl[\log\bar\mu_i-\log\mu(T_i,d_i;\vartheta)\bigr]^{2}}{\sigma_i^{2}}.
\label{eq:objective}
\end{equation}
where $\vartheta$ are the fit parameters, $\hat\vartheta$ their estimated values, $\bar\mu_i$ and $\sigma_i^2$ the mean and variance of each data point. 

We first test $d$-independence of the $T$ decay coefficient, fitting $d=8,16$ separately with $\epsilon(T,d)=a(d) T^{-\gamma}$. The resulting exponent values are $\gamma_8=1.5\pm0.2$ and $\gamma_{16}=1.3\pm0.1$, compatible within errors. 

Consequently, we fit all $d=8,16$ data-points jointly with $\epsilon(T,d) = c d^{\psi\cdot \gamma} T^{-\gamma}$, using two separate parameters for $\mu_\infty(8)$, $\mu_\infty(16)$ but common exponents $\gamma, \psi$. We obtain $\psi=0.06\pm0.21$ and $\gamma=1.4\pm0.1$, suggesting a $d$-independent transient behaviour for $\mu(T,d)$ and determining the limiting values $\mu_\infty(d)$. 

Finally, the large-$d$ extrapolation uses the more stable ratio to the threshold,
$R(d)=\muadv(d)/\mu(d)$ and defines the crossing by $R(\dstar)=1$:
\begin{equation}
\log R(d)=\log R_{0}-\rho\log d,
\qquad
\dstar=R_{0}^{1/\rho}.
\label{eq:Rfit}
\end{equation}
Each point of the $d$-sweep is corrected to its plateau by the fitted transient
\begin{equation}
\tilde\mu_\infty(d)=\frac{\bar\mu(\Tz,d)}{1+c d^{\psi\cdot \gamma} T^{-\gamma}},
\label{eq:fac}
\end{equation}
at $d=2,4,8,16,32$. Because $\psi\simeq0$, the correction is nearly constant in $d$, providing roughly a $26\%$ uniform rescaling. The resulting fit allows to estimate the crossing-value of $d$ for advantage provided in the main text.

\end{document}